\documentclass[a4paper,10pt]{article}

\pdfoutput=1
\usepackage{amsmath}
\usepackage{amssymb}
\usepackage{jcappub}
\usepackage{mathtools}
\usepackage{graphicx}
\usepackage[dvipsnames]{xcolor}
\usepackage{footmisc}
\usepackage[normalem]{ulem}
\usepackage{float}
\usepackage{braket}
\usepackage{wrapfig}
\usepackage{lscape}
\usepackage{rotating}
\usepackage{epstopdf}
\usepackage{comment}
\usepackage[numbers, sort&compress]{natbib}
\usepackage{tikz, pgfplots}
\usetikzlibrary{positioning}
\usetikzlibrary{decorations.pathmorphing}
\usepackage[export]{adjustbox}[2011/08/13]

\tikzset{snake/.style={decorate, decoration=snake}}

\newcommand{\be}{\begin{equation}}
	\newcommand{\ee}{\end{equation}}

\begin{document}
	\title{A semi-analytic estimate for the effective sound speed counterterm in the EFTofLSS}
	\author[1]{Caio Nascimento,}
	\affiliation[1]{Department of Physics, University of Washington, 1410 NE Campus Pkwy, Seattle, USA}
	\emailAdd{caiobsn@uw.edu}
	
	\author[2]{Drew Jamieson,}
	\affiliation[2]{Max-Planck-Institut für Astrophysik, Karl-Schwarzschild-Straße 1, 85748 Garching, Germany}
	\emailAdd{jamieson@mpa-garching.mpg.de}
	
	\author[3]{Matthew McQuinn}
	\affiliation[3]{Department of Astronomy, University of Washington, 1410 NE Campus Pkwy, Seattle, USA
	}
	\emailAdd{mcquinn@uw.edu}

	\author[1]{and Marilena Loverde}
	\emailAdd{mloverde@uw.edu}

\abstract{
The Effective Field Theory of Large Scale Structure (EFTofLSS) has found tremendous success as a perturbative framework for the evolution of large scale structure, and it is now routinely used to compare theoretical predictions against cosmological observations. The model for the total matter field  includes one nuisance parameter at 1-loop order, the effective sound speed, which can be extracted by matching the EFT to full N-body simulations. In this work we first leverage the Layzer-Irvine cosmic energy equation to show that the equation of state can be exactly computed with knowledge of the fully nonlinear power spectrum. When augmented with separate universe methods, we show one can estimate  the effective sound speed. This estimate is in good agreement with simulation results, with errors at the few tens of percent level. We apply our method to investigate the cosmology dependence of the effective sound speed and to shed light on what cosmic structures shape its value.}

\maketitle

\section{Introduction}
\label{sec:int}

In the past fifteen years the Effective Field Theory of Large Scale Structure (EFTofLSS) \cite{Baumann:2010tm, Carrasco:2012cv, Carroll:2013oxa, Pietroni:2011iz, Carrasco:2013mua, Porto:2013qua} has proved itself to be a very powerful framework, enabling one to push the regime of validity of standard perturbation theory (SPT) methods \cite{Makino:1991rp, Jain:1993jh, Goroff:1986ep, Zeldovich:1969sb, Bernardeau:2001qr, Scoccimarro:1995if} for the large-scale structure evolution to smaller scales \cite{Baldauf:2016sjb, Braganca:2023pcp, Nishimichi:2020tvu}. The EFTofLSS is now an integral part of theory modeling involved in the analysis of real data \cite{Euclid:2023tog,Chudaykin:2020aoj, Ivanov:2019pdj,Philcox:2022frc, Taule:2023izt, Beutler:2011hx, Blake:2011rj, Chudaykin:2022nru, Taule:2024bot, Aviles:2024zlw}.

While cosmological simulations are strictly necessary to ensure accuracy of modeling on sufficiently small scales \cite{Carlson:2009it} where the nonlinear gravitational evolution becomes strongly coupled, perturbative methods offer a number of advantages and are thus complementary to simulations. They provide the kind of intuition that only an analytic model can offer, with flexibility to go beyond standard scenarios \cite{Bottaro:2023wkd, Piga:2022mge, Lewandowski:2019txi, Crisostomi:2019vhj, Bottaro:2024pcb}, and circumvent the need for extensive computational resources that simulations require \cite{Schneider:2015yka}. Additionally, semi-analytic methods can provide insights on the nonperturbative regime of nonlinear structure formation as well (e.g. \cite{Gunn:1972sv, Press:1973iz, Bond:1990iw, Ivanov:2018lcg, Berlind:2001xk}).

The EFTofLSS was created to fix two major shortcomings of SPT: To accommodate deviations from the assumption of a pressureless fluid \footnote{Even collisionless systems have a nonzero pressure induced by short-wavelength fluctuations. This is exactly the subject of study in this manuscript.} and to address the issue of sensitivity to uncontrolled short distance physics \footnote{As can be seen from the fact that modes with arbitrarily high frequencies are running on the loops in SPT.}. At one-loop, this comes at the cost of introducing a new free parameter, the effective sound speed counterterm, which is to be determined either by observations or through a matching to N-body simulations. The one-loop power spectrum in the EFT reads
\begin{equation}
\label{eq:powerresintro}
	P_{\textrm{1-loop}}(a,k; \Lambda) =  P_{\textrm{1-loop,SPT}}(a,k; \Lambda) -D_{\textrm{L}}(a)c_{\textrm{ctr}}(a;\Lambda)k^2 W_{\Lambda}^{2}(k) P_{\textrm{L}}(k) \,,
\end{equation}
where $P_{\textrm{1-loop,SPT}}(a,k; \Lambda)$ is the one-loop power spectrum in SPT, $D_{\textrm{L}}(a)$ is the linear growth factor, $P_{\textrm{L}}(k)$ is the linear theory power spectrum, $W_{\Lambda}(k)$ is a window function introduced to smooth the density field over some distance scale $\Lambda^{-1}$, and $c_{\textrm{ctr}}(a;\Lambda)$ is the counterterm [which is directly related to the effective sound speed of dark matter when treated as a fluid, $c_{\textrm{eff}}^{2}(a;\Lambda)$, as detailed in Appendix \ref{sec:app2}, see Eq.~(\ref{eq:counterapp})]. One eventually takes the $\Lambda \to \infty$ limit, which removes the sensitivity to a choice of window function. 

In this paper we focus on the effective sound speed parameter, which enters in all applications of the EFTofLSS. However, additional free parameters become necessary when considering higher order terms in the perturbative expansion \cite{Konstandin:2019bay}, when modeling biased tracers \cite{Desjacques:2016bnm} and also for higher point correlation functions (at the same loop precision) \cite{Baldauf:2014qfa, Baldauf:2021zlt, Bertolini:2016bmt}. We will consider the simplest case of one-loop perturbation theory to describe the two-point clustering of matter. This is sufficient as a perturbative description of weak gravitational lensing on sufficiently large scales, but not of galaxy clustering which also requires modeling of galaxy formation physics (even on large scales via a bias expansion). We will also work in real space, i.e., ignoring redshift space distortions \cite{Senatore:2014vja}.

The effective sound speed is expected to account for the backreaction on large scales of nonlinear effects associated to gravitational collapse and the subsequent formation of the cosmic web in the form of cosmic sheets, filaments and halos \cite{Pueblas:2008uv, Buehlmann:2018qmm, Ivanov:2022mrd, Kugel:2024zxq}. The counterterm has been successfully extracted from matching EFT predictions to simulations in a number of previous works \cite{Carrasco:2012cv,Foreman:2015lca, Baldauf:2015aha,Carrasco:2013sva, Karandikar:2023ozp,Baldauf:2015tla,McQuinn:2015tva, Ivanov:2018lcg, Senatore:2014via, Angulo:2014tfa, Foreman:2015uva}, but such measurements do not offer any clear physical interpretation for the underlying microphysics responsible for the effective sound speed. In particular, a simple estimate of the numerical value of that quantity from some semi-analytic model for the short scale fluctuations is still lacking (beyond just dimensional analysis, e.g. see \cite{Ivanov:2022mrd}). This illustrates the point that the microphysics associated to the EFT counterterm is yet to be fully understood. Such an understanding would shed light on the cosmology dependence of the sound speed parameter, and on what cosmic structures shape its value.

In this work we will make progress in that direction by providing a simple semi-analytic estimate for the effective sound speed, which we show is in very good agreement with simulation results. We first show that we can leverage the Layzer-Irvine cosmic energy equation to exactly compute the equation of state, i.e. the ratio between background pressure and energy density, assuming knowledge of the fully nonlinear matter power spectrum. Since the sound speed is related to the response of the equation of state to the presence of a long-wavelength mode, this introduces the necessary ingredients to estimate the effective sound speed using separate universe techniques. The estimate reproduces the results from high resolution simulations with errors at the tens of percent level.

We will work with a $\Lambda$CDM universe, particularly its late time dynamics (redshifts $z \lesssim  100$) on subhorizon scales $k \gg aH$ [where $a(t)$ is the cosmological scale factor, $H=d\log a/dt$ is the Hubble expansion rate and $t$ is cosmic time]. In this regime one can write Newtonian equations of motion in an expanding cosmological background, and the collective dynamics of cold dark matter particles is encoded in the collisionless Boltzmann equation coupled to the Poisson equation, the so-called Vlasov-Poisson system. 

Unless explicitly stated otherwise, for numerical calculations we assume a fiducial cosmology $\Lambda CDM_{\textrm{DarkSky}}$ with $\Omega_{\textrm{m}} =0.295$, $\Omega_{\Lambda}=0.705$, $h=0.688$, $n_{s}=0.9676$ and $\sigma_{8}=0.835$, matching the choice made in \cite{Foreman:2015lca}, which adopts the Dark Sky simulations \cite{Skillman:2014qca}. All power spectra are computed using the Boltzmann solver CLASS \cite{blas2011cosmic}, in combination with HMcode \cite{Mead:2020vgs} for the nonlinear power spectrum when needed (without baryonic feedback effects), at $16$ equally spaced redshift values in the interval $0 \leq z \leq 5$. We interpolate in-between and for $z >5$ assume a linear-theory evolution in all cases, i.e., $P(z,k) = [D_{\textrm{L}}(z)/D_{\textrm{L}}(z=5)]^2 P(z=5,k)$, where $D_{\textrm{L}}(a)$ is the linear growth factor (we use the fitting formula provided in \cite{Hamilton:2000tk}) \footnote{HMcode is calibrated against simulations at $z \leq 2$. In that redshift range, it is accurate out to $k=10h\textrm{Mpc}^{-1}$ at the few percent level \cite{Mead:2020vgs}. We also used the EuclidEmulator2 (EE2), which is accurate at the percent level in the same range of scales and redshifts \cite{Euclid:2020rfv}, to cross-check our results. We obtained nearly identical results using either HMcode or EE2.}. Although we only numerically compute quantities of interest at $z \lesssim 2$, integrals over the scale factor appear that require the nonlinear power spectrum out to arbitrarily high redshift. However, we have checked that when computing the equation of state and the effective sound speed, contributions from $z>5$ are essentially negligible, which justifies our simple linear-theory scaling. 

The paper is organized as follows: In Sec.\ref{sec:eftoflss} we review the standard EFTofLSS framework to set notation and collect some important results. In Sec.\ref{sec:eos} we use the Layzer-Irvine equation to compute the equation of state assuming knowledge of the fully nonlinear power spectrum. In Sec.\ref{sec:eff_cs} we apply the tools developed in the previous section, in combination with separate universe methods, to derive an estimate for the effective sound speed which we compare to simulation results. In Sec.\ref{sec:concl} we summarize our findings. Appendices \ref{sec:app1}, \ref{sec:app2} and \ref{sec:app3} supplement the main text with relevant discussions and derivations.

\section{EFTofLSS framework}
\label{sec:eftoflss}

In this section we review the EFTofLSS framework, mostly following the phase-space approach of \cite{Carrasco:2012cv}, to establish notation and recollect some well-known results that will be used in later sections. 

The collective behavior of particles interacting only via the gravitational force in an expanding universe is governed by the collisionless Boltzmann (or Vlasov) equation:
\begin{equation}
	\label{eq:vlasov}
	\frac{\partial f}{\partial \eta} + \frac{d\vec{x}}{d\eta} \cdot \frac{\partial f}{\partial \vec{x}} + \frac{d\vec{q}}{d\eta} \cdot \frac{\partial f}{\partial \vec{q}} = 0 \, ,
\end{equation}
for the phase space distribution function $f(\eta, \vec{x},\vec{q})$. We work with the superconformal time defined by $d\eta = dt/a^2(t)$ (sometimes we also use the scale factor as the time variable). Additionally, $\vec{x}$ are comoving coordinates and $\vec{q}$ is the comoving momentum such that $(d\vec{x}/d\eta) = a^2 (d\vec{x}/dt) = (\vec{q}/m) \equiv \vec{q}$, with $m$ the cold dark matter (CDM) particle mass which we set to unity from this point forward. We then have $(d\vec{q}/d\eta) = -ma^2 \vec{\nabla} \phi = - a^2 \vec{\nabla} \phi$, with $\phi(\eta,\vec{x})$ the Newtonian gravitational potential. Eq.~(\ref{eq:vlasov}) now reads,
\begin{equation}
	\label{eq:vlasov2}
	\frac{\partial f}{\partial \eta} + \vec{q} \cdot \frac{\partial f}{\partial \vec{x}} = a^2(\eta) \frac{\partial \phi}{\partial \vec{x}} \cdot \frac{\partial f}{\partial \vec{q}} \,.
\end{equation}
The gravitational potential is not independent from the distribution function, but rather determined by the Poisson equation
\begin{equation}
	\label{eq:poisson}
	\nabla^2 \phi = 4\pi Ga^2 (\rho-\bar{\rho}) \,,
\end{equation}
where
\begin{equation}
	\label{eq:energydensity}
	\rho(\eta, \vec{x}) = a^{-3}(\eta) \int \frac{d^3 \vec{q}}{(2\pi)^{3}} f(\eta,\vec{x},\vec{q}) \,,
\end{equation}
is the energy density. \footnote{In our convention of unity CDM particle mass, this becomes the number density. Similar statements can be made on the other fluid quantities as well.} Note that only the fluctuations around the (ensemble) average density, $\bar{\rho}(\eta) = \langle \rho(\eta,\vec{x}) \rangle \propto a^{-3}(\eta)$, contribute to the gravitational potential.

The nonlinear term on the right-hand side of Eq.~(\ref{eq:vlasov2}) is a contact term, \footnote{By contact term we mean a product of two fields evaluated at the same point in space.} being sensitive to nonlinearities in gravitational dynamics at arbitrarily small scales. In order to have theoretical control over the backreaction from the uncertain short scale modes, we follow the standard EFT approach and split the distribution function into a long-wavelength piece defined by smoothing over some (distance) scale $1/\Lambda$
\begin{equation}
	\label{eq:smoothing}
	f_{l}(\eta,\vec{x},\vec{q}) = \int d^{3} \vec{x}' \, W_{\Lambda}(|\vec{x}-\vec{x}'|)f(\eta,\vec{x}',\vec{q}) \,,
\end{equation}
and a short-wavelength piece $f_{s}=f-f_{l}$, via a window function $W_{\Lambda}(x)$ \footnote{For numerical calculations we follow the standard convention in the EFTofLSS literature and consider a Gaussian filter, i.e., $W_{\Lambda}(k) = e^{-k^2/2\Lambda^2}$ in Fourier space. We assume a time-independent filter. One could think of working with a time-dependent filter, since more modes are perturbative at higher redshift, as follows from the time-dependence of the scale of nonlinearities. However, this adds unnecessary technical complications to the model since the ultimate goal is to take the $\Lambda \to \infty$ limit.}. A similar decomposition, $\phi = \phi_{l}+\phi_{s}$, also follows for the gravitational potential since it is linearly related to the distribution function by Eqs.~(\ref{eq:poisson}) and (\ref{eq:energydensity}). 

The evolution equation for the long-wavelength piece now follows from Eqs.~(\ref{eq:vlasov2}) and (\ref{eq:smoothing})
\begin{equation}
	\label{eq:vlasovlong}
	\frac{\partial f_{l}}{\partial \eta} + \vec{q} \cdot \frac{\partial f_{l}}{\partial \vec{x}} = a^{2}(\eta) \int d^{3}\vec{x}' \, W_{\Lambda}(|\vec{x}-\vec{x}'|) \frac{\partial \phi (\eta,\vec{x}')}{\partial \vec{x}'} \cdot \frac{\partial f(\eta,\vec{x}',\vec{q})}{\partial \vec{q}} \,.
\end{equation}
Following standard methods (e.g., see \cite{Baumann:2010tm}) one obtains the result that mixed terms involving a product of long and short modes are suppressed, and hence
\begin{equation}
	\label{eq:vlasovlong2}
	\frac{\partial f_{l}}{\partial \eta} + \vec{q} \cdot \frac{\partial f_{l}}{\partial \vec{x}} = a^{2}(\eta) \frac{\partial \phi_{l}}{\partial \vec{x}} \cdot \frac{\partial f_{l}}{\partial \vec{q}} + a^2(\eta) \left[\frac{\partial \phi_{s}}{\partial \vec{x}} \cdot \frac{\partial f_{s}}{\partial \vec{q}} \right]_{\Lambda} + O\Big(\frac{k^2}{\Lambda^2}\Big) \,,
\end{equation}
where,
\begin{equation}
	\label{eq:shortmodes}
	\left[\frac{\partial \phi_{s}}{\partial \vec{x}} \cdot \frac{\partial f_{s}}{\partial \vec{q}} \right]_{\Lambda} = \int d^3 \vec{x}' \, W_{\Lambda}(|\vec{x}-\vec{x}'|) \frac{\partial \phi_{s} (\eta,\vec{x}')}{\partial \vec{x}'} \cdot \frac{\partial f_{s}(\eta,\vec{x}',\vec{q})}{\partial \vec{q}} \,.
\end{equation}
From this point forward we will drop the higher derivative corrections of order $\mathcal{O}(k^2/\Lambda^2)$. They vanish in the $\Lambda \to \infty$ limit, and do not contribute to the effective sound speed even at finite $\Lambda$. However, it is important to keep in mind that their presence become necessary to accurately model the nonlinear power spectrum for any finite value of the cutoff \cite{Baumann:2010tm, Carrasco:2012cv}. In the remainder of this section we delve into the effective stress tensor in the EFTofLSS.

\subsection{Effective stress tensor}
\label{sec:effstress}

\begin{figure}
	\centering
	\includegraphics[width=0.75\textwidth]{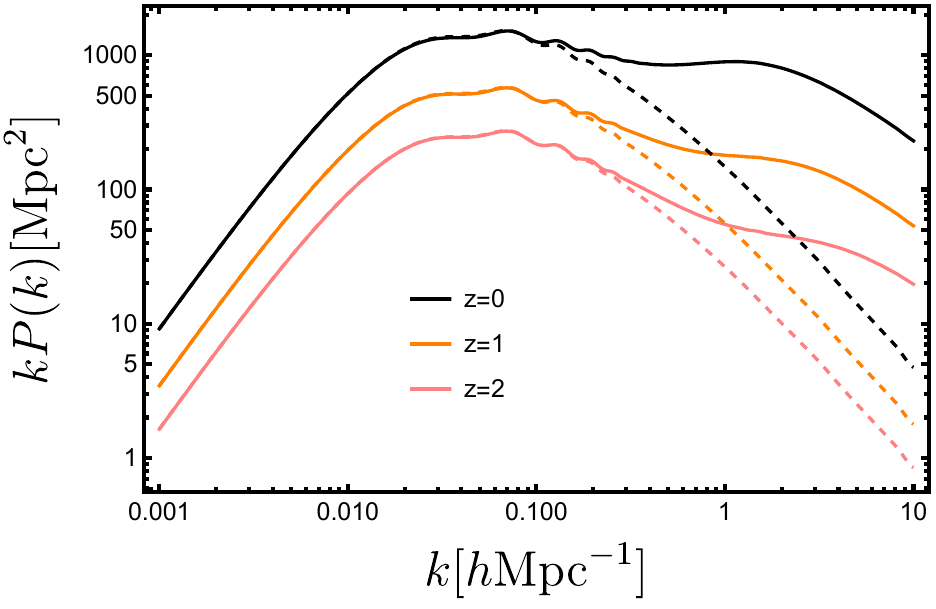}
	\caption{Plot of $kP(k)$, which is integrated over $\log k$ in Eq.~(\ref{eq:binding_calculated}), as a function of the comoving wavenumber $k$. Solid lines show the fully nonlinear matter power spectrum and dashed lines correspond to the linear theory matter power spectrum. Note the presence of a second peak at $k\sim 1h\textrm{Mpc}^{-1}$, which causes the short-scale gravitational binding energy in Eq.~(\ref{eq:binding_calculated}) to pick up sizable contributions from very small scales of a few $\sim h\textrm{Mpc}^{-1}$, typical of virialized halos.}
	\label{fig:mps}
\end{figure}

By taking the first two moments of Eq.~(\ref{eq:vlasovlong2}), one can derive the dynamical equations for the long-wavelength effective fluid. This calculation can be found in the literature \cite{Carrasco:2012cv, McQuinn:2015tva} and leads to the continuity and Euler equations. The former corresponds to conservation of mass, and reads
\begin{equation}
	\label{eq:continuity}
	\rho_{l}' + 3\mathcal{H} \rho_{l} +a \vec{\nabla} \cdot \vec{\Pi}_{l} = 0 \, ,
\end{equation}
where prime denotes a derivative with respect to superconformal time, $\mathcal{H} = d\log a/d\eta = a^2H$, and
\begin{equation}
	\label{eq:momentum}
	\vec{\Pi}_{l}(\eta,\vec{x}) = a^{-4}(\eta) \int \frac{d^3 \vec{q}}{(2\pi)^{3}} \, \vec{q} \, f_{l}(\eta,\vec{x},\vec{q}) \,, 
\end{equation}
is the fluid momentum. The Euler equation is
\begin{equation}
\label{eq:euler2}
	v_{l,i}' + \mathcal{H} v_{l,i} +a v_{l,j} \partial^{j} v_{l,i} + a \partial_{i} \phi_{l} + \frac{a}{\rho_{l}} \partial^{j} \tau^{\textrm{eff}}_{ij} = 0 \,,
\end{equation}
where
\begin{equation}
	\label{eq:velocity}
	\vec{\Pi}_{l}(\eta,\vec{x}) = \rho_{l}(\eta,\vec{x}) \vec{v}_{l}(\eta,\vec{x})\,, 
\end{equation}
defines the long-wavelength velocity field $\vec{v}_{l}(\eta,\vec{x})$. The effective stress tensor is a sum of kinetic and potential terms:
\begin{equation}
	\label{eq:effstress}
	\tau^{\textrm{eff}}_{ij}(\eta,\vec{x}) = 2K_{ij}(\eta,\vec{x}) + U_{ij}(\eta,\vec{x}) \,.
\end{equation}
The kinetic part is
\begin{equation}
\begin{split}
\label{eq:kinetic}
	 K_{ij}(\eta,\vec{x}) & = \frac{1}{2} a^{-5}(\eta) \int \frac{d^3 \vec{q}}{(2\pi)^3}  f_{l}(\eta,\vec{x},\vec{q}) \left[ q_{i} - a(\eta)v_{l i}(\eta, \vec{x})\right] \left[ q_{j} - a(\eta)v_{l j}(\eta, \vec{x})\right] \\ & = \frac{1}{2} \left[ a^{-5}(\eta) \int \frac{d^3 \vec{q}}{(2\pi)^3} f_{l}(\eta,\vec{x},\vec{q}) q_{i}q_{j} - \rho_{l}(\eta,\vec{x})v_{li}(\eta,\vec{x})v_{lj}(\eta,\vec{x}) \right] \,,  
\end{split}
\end{equation}
where the second equality follows from Eqs.~(\ref{eq:momentum}) and (\ref{eq:velocity}). The potential part reads
\begin{equation}
\label{eq:potential}
	U_{ij}(\eta,\vec{x}) = \frac{1}{4\pi G a^2(\eta)} \left\{  \left[ \partial_{i} \phi_{s} \partial_{j} \phi_{s} \right]_{\Lambda}  -\frac{1}{2} \left[\partial_{k} \phi_{s} \partial^{k} \phi_{s}\right]_{\Lambda} \delta_{ij} \right\}  \,. 
\end{equation}
The kinetic contribution in Eq.~(\ref{eq:kinetic}) is sourced by the velocity dispersion while the potential contribution in Eq.~(\ref{eq:potential}) derives from the short scale gravitational potential. Let us introduce some notation for the average of these quantities. The short-scale kinetic energy per unit mass, $\kappa(\eta)$, is defined as
\begin{equation}
	\label{eq:kappa}
	\bar{\rho}(\eta) \kappa (\eta) \equiv \langle K^{i}_{i}(\eta,\vec{x}) \rangle \,.
\end{equation}
Similarly the short-scale gravitational binding energy per unit mass, $u(\eta)$, is defined as
\begin{equation}
	\label{eq:Omega}
	\bar{\rho}(\eta) u(\eta) \equiv \langle U^{i}_{i}(\eta,\vec{x}) \rangle \,,
\end{equation}
in terms of ensemble averages of kinetic and potential contributions to the effective stress. The traceless parts of $K_{ij}(\eta,\vec{x})$ and $U_{ij}(\eta,\vec{x})$ vanish upon averaging due to statistical isotropy. Note from the second line in Eq.~(\ref{eq:kinetic}), as well as Eq.~(\ref{eq:kappa}), that the kinetic energy from the bulk flow is subtracted from the total kinetic energy in the definition of $\kappa(\eta)$, with only the contribution from velocity dispersion remaining, hence we refer to it as the short-scale kinetic energy per unit mass. Similarly, from Eqs.~(\ref{eq:potential}) and (\ref{eq:Omega}) it follows that
\begin{equation}
	\label{eq:binding}
	\bar{\rho} u = - \frac{1}{8\pi Ga^2} \langle \vec{\nabla} \phi_{s} \cdot \vec{\nabla} \phi_{s} \rangle = \frac{1}{2} \langle \phi_{s} \rho_{s} \rangle \implies u = \frac{1}{2} \langle \phi_{s} \delta_{s} \rangle \,,
\end{equation}
where we integrate by parts inside the ensemble average \footnote{The ensemble average can always be thought of as a volume average, which involves integrating over position. This justifies an integration by parts under the average sign. Also note that since these are global averages, the integral over positions are defined over the entire volume and hence boundary terms can be assumed negligible with appropriate boundary conditions at spatial infinity.  \label{footnote2}}, using the Poisson Eq.~(\ref{eq:poisson}), and from now on we often omit time and scale dependencies to simplify the notation. This shows that $u(\eta)$ is indeed the short-scale gravitational binding energy. Before showing how to compute this quantity we first introduce the Friedmann equation:
\begin{equation}
	\label{eq:fried}
	H^2 = \frac{8\pi G}{3} \rho_{\textrm{cri}} = \frac{8\pi G}{3} \frac{\bar{\rho}}{\Omega_{\textrm{m}}(a)} \,,
\end{equation}
where $\Omega_{\textrm{m}}(a) = \bar{\rho}(a)/\rho_{\textrm{cri}}(a)$ is the fractional contribution of matter to the total (critical) energy density. Evaluating Eq.~(\ref{eq:fried}) at the present time, and combining it with Eq.~(\ref{eq:poisson}), allows us to write the Poisson equation in the following form in Fourier space ($\vec{\nabla} \to i\vec{k}$)
\begin{equation}
	\label{eq:poisson2}
	k^2 \phi = -\frac{3}{2} \Omega_{\textrm{m},0}H_{0}^{2} \frac{\delta}{a} \,,
\end{equation}
where we used $\bar{\rho} \propto a^{-3}$, and quantities evaluated at the present time carry a subscript $0$, such as the Hubble expansion rate today $H_{0} = 100h(\textrm{km/s})/\textrm{Mpc}$. We can now substitute Eq.~(\ref{eq:poisson2}) into Eq.~(\ref{eq:binding}) to arrive at
\begin{equation}
	\label{eq:binding_calculated}
	u(a) = -\frac{3}{4} \Omega_{\textrm{m},0} H_{0}^2 \frac{1}{a} \int \frac{dk}{2\pi^2} P(a,k) \left[1-W_{\Lambda}(k)\right]^2 \,,
\end{equation}
where
\begin{equation}
	\label{eq:powerspectrum}
	\langle \delta(a,\vec{k}) \delta(a,\vec{k'}) \rangle = (2\pi)^3\delta^{(3)}(\vec{k}+\vec{k'})P(a,k) \,,
\end{equation}
defines the fully nonlinear matter power spectrum $P(a,k)$, which Eq.~(\ref{eq:binding_calculated}) shows is all one needs to compute the short-scale gravitational binding energy per unit mass. In Fig.~\ref{fig:mps} we plot $kP(k)$, which is integrated over $\log k$ in Eq.~(\ref{eq:binding_calculated}), for a few redshift values. Solid and dashed lines correspond to the fully nonlinear and linear theory power spectra, respectively. It is clear that at $z=0$, $kP(k)$ has a second peak at a small scale of $k \sim 2h\textrm{Mpc}^{-1}$ in the nonlinear case. This second peak is less pronounced at higher redshift and eventually disappears as the linear and nonlinear power spectra start to converge toward one another. For that reason, the integral in Eq.~(\ref{eq:binding_calculated}) picks up significant contributions from scales of a few $\sim h\textrm{Mpc}^{-1}$ at low redshift. This shows that the gravitational binding energy picks-up large contributions from virialized halos.  

\begin{figure}
	\centering
	\includegraphics[width=0.75\textwidth]{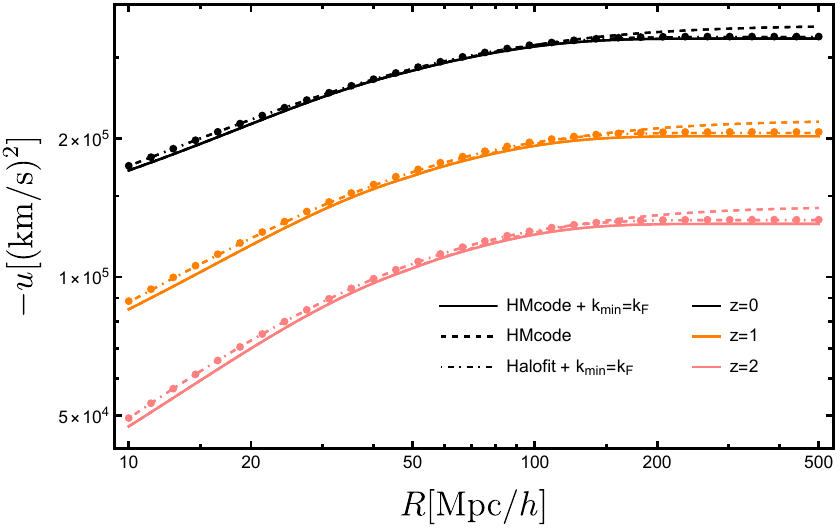}
	\caption{Short-scale gravitational binding energy per unit mass as a function of the smoothing scale $R=\Lambda^{-1}$, for varying redshift. Round data points correspond to results from the dark-matter-only runs of the MilleniumTNG simulations \cite{Hernandez-Aguayo:2022xcl}. Solid, dashed and dot-dashed lines correspond to different theory predictions from Eq.~(\ref{eq:binding_calculated}), differing on the choice of code used to model the nonlinear power spectrum and whether or not a hard cutoff is introduced at the fundamental mode of the simulation (as indicated in the legends). There is excellent agreement between theory and simulation results, with Halofit performing slightly better than HMcode.  }
	\label{fig:pot}
\end{figure}

In Fig.~\ref{fig:pot} we plot the short-scale gravitational binding energy as a function of the smoothing scale $R=\Lambda^{-1}$, for a few redshift values. The round data points were extracted from averaging the two phase-reversed runs of the dark-matter-only MilleniumTNG simulations \cite{Hernandez-Aguayo:2022xcl}. The simulations contain $2160^3$ dark matter particles in a $L_{\textrm{box}} = 500~\rm{Mpc}~h^{-1}$ box \footnote{We compute the binding energies from the simulation snapshots using an Eulerian density mesh on high-resolution 3D grid of $2048^3$ voxels, constructed using the cubic spline assignment scheme. We Fourier transform the density field, solve the Poisson equation for the modes of the potential, and inverse Fourier transform the potential modes back to real space. Then we smooth both the original density field and the Newtonian potential with Gaussian filters. By subtracting the smoothed fields from the original fields, we can compute the mean binding energy by averaging over the grid, using Eq.~(\ref{eq:binding}). }. Since we do not have access to many realizations of the simulation, we do not include errors bars due to cosmic variance (which is a source of uncertainty on the largest scales, but is mitigated by averaging over the two phase-reversed runs). Solid lines are the theory predictions, from Eq.~(\ref{eq:binding_calculated}), using HMcode to compute the nonlinear power spectrum, and also adding a hard infrared cutoff at the fundamental mode of the simulation $k_{\textrm{F}}=2\pi/L_{\textrm{box}}$. The dashed line is the theory prediction integrated over all wavenumber, and the dot-dashed line uses Halofit \cite{Takahashi:2012em} to compute the nonlinear power spectrum, and also adds a hard cutoff at the fundamental mode. The theory and simulation results are in excellent agreement, with Halofit perfoming slightly better than HMcode.

We will see next that the short-scale kinetic energy can be computed from $u(a)$, which is hence the basic ingredient needed in the calculation of the average effective stress tensor, and also of the effective sound speed as we will see in Sec.\ref{sec:eff_cs}.

\subsection{Cosmic energy equation}
\label{eq:liceq}

In the previous section we established that the short-scale gravitational binding energy can be straightforwardly calculated assuming knowledge of the fully nonlinear power spectrum. On the other hand, one could \textit{apriori} suspect that the short-scale kinetic energy is much harder to extract and requires simulations. However, by using energy conservation arguments we can expect to relate the kinetic and potential terms \cite{Siegel:2005xu, Chiang:2020ssj}. This is achieved by the Layzer-Irvine cosmic energy equation \cite{peebles1980large, Baumann:2010tm}:
\begin{equation}
	\label{eq:cosmic_energy}
	\frac{d}{d\eta} (\kappa + u) + \mathcal{H} (2\kappa + u) = 0 \,,
\end{equation}
which generalizes the concept of energy conservation to an expanding space that breaks time-translation symmetry, so that energy is not conserved in general. From Eq.~(\ref{eq:cosmic_energy}) we see that energy conservation applies only for virialized scales for which $2\kappa+u=0$. 

Let us also emphasize that both $\kappa$ and $u$ appearing in Eq.~(\ref{eq:cosmic_energy}) are short-scale quantities as defined in Eqs.~(\ref{eq:kappa}) and (\ref{eq:Omega}), while the Layzer-Irvine equation is often phrased in terms of the total kinetic and potential energies with contributions from all scales. We present a derivation of Eq.~(\ref{eq:cosmic_energy}) from the Boltzmann equation in Appendix \ref{sec:app1}, which holds true because contributions from long-wavelengths separately satisfy the very same equation, and can hence be subtracted.

We will be primarily interested in the integral form of Eq.~(\ref{eq:cosmic_energy}) \footnote{We drop integration constants since they can be made negligible by choosing the initial time to be sufficiently early, and simply set $\eta_{i}=0$. The resulting integral is dominated by its upper limit.}, 
\begin{equation}
	\label{eq:cosmic_energy_int}
	\kappa(\eta) = -a^{-2}(\eta) \int_{0}^{\eta} d\eta' a(\eta') \frac{d}{d\eta'}\left[a(\eta')u(\eta')\right] \,.
\end{equation}
This formula shows that a nonzero average velocity dispersion is effectively sourced by a short-scale gravitational binding energy, and is hence a short-wavelength quantity (also see Appendix A in \cite{Carrasco:2012cv}).

Integrating Eq.~(\ref{eq:cosmic_energy_int}) by parts leads to
\begin{equation}
	\label{eq:kit_int}
	\kappa(a) = -u(a) +a^{-2} \int_{0}^{a} da' a' u(a') \,,
\end{equation}
where we now use the scale factor as our clock. This can be applied to compute the short-scale kinetic energy per unit mass from Eq.~(\ref{eq:binding_calculated}), and hence also the root mean square velocity dispersion $\sigma_{\textrm{dis}} = \sqrt{2\kappa}$. 

At this point it is instructive to see how Eq.~(\ref{eq:kit_int}) simplifies in the case where we use the linear theory power spectrum, $P(a,k) = D_{\textrm{L}}^2(a)P_{\textrm{L}}(k)$, to evaluate the gravitational binding energy in Eq.~(\ref{eq:binding_calculated}). In this case, Eq.~(\ref{eq:binding_calculated}) becomes
\begin{equation}
	\label{eq:binding_calculated_linear}
	u = -\frac{3}{4} \Omega_{\textrm{m},0} H_{0}^2 \frac{D_{\textrm{L}}^2}{a} \int \frac{dk}{2\pi^2} P_{\textrm{L}}(k) \left[1-W_{\Lambda}(k)\right]^2 \,.
\end{equation}
Substituting Eq.~(\ref{eq:binding_calculated_linear}) into Eq.~(\ref{eq:cosmic_energy_int}) yields
\begin{equation}
\label{eq:linearkappa}
\begin{split}
  \kappa & = \int \frac{dk}{2\pi^2} P_{\textrm{L}}(k) \left[1-W_{\Lambda}(k)\right]^2 \times \frac{3}{2} \Omega_{\textrm{m},0} H_{0}^2 a^{-2} \int_{0}^{\eta} d\eta' a(\eta') D_{\textrm{L}}(\eta') \frac{dD_{\textrm{L}}}{d\eta'} \\ & = \frac{1}{2} a^2 H^2 f^2 D_{\textrm{L}}^2 \int \frac{dk}{2\pi^2} P_{\textrm{L}}(k) \left[1-W_{\Lambda}(k)\right]^2 \,,
\end{split}
\end{equation}
where we introduced the linear growth rate $f(a)= d\log D_{\textrm{L}}/d\log a$, and to arrive at the second line in Eq.~(\ref{eq:linearkappa}) we used the differential equation satisfied by the linear growth factor
\begin{equation}
\label{fig:difflg}
	\frac{d^2 D_{\textrm{L}}}{d\eta^2} = \frac{3}{2} \Omega_{\textrm{m},0} H_{0}^2 a(\eta) D_{\textrm{L}}(\eta) \,,
\end{equation}
to do the remaining time integral in the first line of Eq.~(\ref{eq:linearkappa}). The final formula Eq.~(\ref{eq:linearkappa}) corresponds to the linear theory velocity field integrated over short scales $k>\Lambda$, precisely as one would expect. 

\begin{figure}
	\centering
	\includegraphics[width=0.75\textwidth]{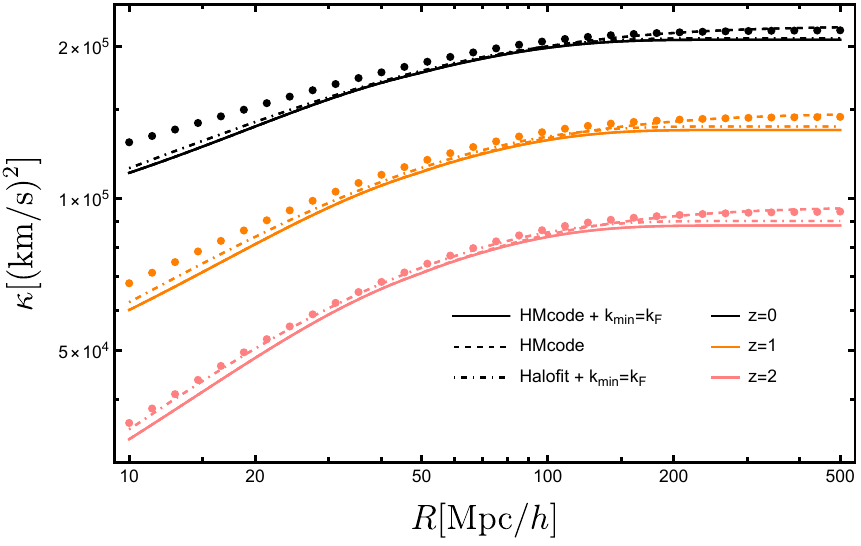}
	\caption{Short-scale kinetic energy per unit mass as a function of the smoothing scale $R=\Lambda^{-1}$, for varying redshift. Round data points correspond to results from the dark-matter-only runs of the MilleniumTNG simulations \cite{Hernandez-Aguayo:2022xcl}. Solid, dashed and dot-dashed lines correspond to different theory predictions based on the Lyzer-Irvine Eq.~(\ref{eq:kit_int}), differing on the choice of code used to model the nonlinear power spectrum and whether or not a hard cutoff is introduced at the fundamental mode of the simulation (as indicated in the legends). There is good agreement between theory and simulation results, with Halofit performing slightly better than HMcode (see discussion on the main text for comments on the small discrepancies). This confirms the validity of the cosmic energy equation. }
	\label{fig:vel_dis}
\end{figure}

To verify the validity of the cosmic energy equation in the nonlinear regime we extract the root mean square velocity dispersion, smoothed on scales $R=\Lambda^{-1}$, from averaging the two phase-reserved runs of the dark-matter-only MilleniumTNG simulations (something similar was previously done in \cite{VIRGOConsortium:1997rhx} as well). In Fig.~\ref{fig:vel_dis} we plot the short-scale kinetic energy as a function of the smoothing scale $R=\Lambda^{-1}$, for varying redshift. Simulation results are shown as round data points \footnote{We computed the N-body velocity dispersion by taking the particle-wise root-mean-square (RMS) of the simulation velocities. To subtract the long-range contribution, we first estimate the momentum field by distributing the particle velocities to a high-resolution 3D grid of $2048^3$ voxels using the cubic spline assignment scheme. We also construct the density field the same way. We then divide the momentum field by the density field to obtain the velocities on the grid. After applying a Gaussian filter of radius $R$, we interpolate the smoothed velocities back to the N-body particles and subtract the smoothed velocities from the original particle velocities. Note, the particle-wise average is automatically density weighted. Finally, the short-scale kinetic energy follows from $\kappa = \sigma_{\textrm{dis}}^2/2$.}. Solid lines correspond to the theoretical expectation based on the Layzer-Irvine Eq.~(\ref{eq:kit_int}), computed using HMcode for the nonlinear power spectrum and also adding a hard cutoff at the fundamental mode of the simulation $k_{\textrm{F}}=2\pi/L_{\textrm{box}}$. The dashed line is the theory prediction integrated over all wavenumber, and the dot-dashed line uses Halofit to compute the nonlinear power spectrum, and also adds a hard cutoff at the fundamental mode. 

There is good agreement between theory and simulation results, confirming the validity of the cosmic energy Eq.~(\ref{eq:kit_int}). Once again Halofit perfoms slightly better than HMcode, but there are residual differences to the simulation results, which prefer slightly larger velocity dispersion, and increase on the smaller scales at later times. We were unable to pinpoint what is causing the discrepancies. Shot noise would increase the simulation measurements, but this leads to the question of why velocities are more sensitive to the shot noise than the densities, and we would expect it to be worse at early times. Another potential source of systematics are empty cells, which are problematic since we need to divide by density to get the velocities \footnote{This is dealt with by averaging an empty cell with all of its nearest neighbors. This smooths out the density and velocity fields only for these underdense cells. This could potentially slightly over-predict their velocities, and the effect would be more significant at later times and at smaller scales since a stronger clustering implies a larger abundance of empty cells. }. It is important to emphasize that cosmic variance cannot explain any discrepancies on the largest scales, since Fig.\ref{fig:pot} shows excellent agreement even on such scales, and the actual realization of the kinetic energy (i.e. the round data points) can be reconstructed exactly from the actual realization of the gravitational energy.

\section{The equation of state}
\label{sec:eos}

In Sec.\ref{sec:effstress} we reviewed that short-wavelength fluctuations, which are not under perturbative control, contribute to the dynamics of long-wavelength fluctuations via an effective stress tensor. In this section we compute its ensemble average, which is fully specified by the pressure due to statistical isotropy. After dividing by the background density this produces the equation of state, $\omega(a)$, defined as follows
\begin{equation}
	\label{eq:eos}
	\omega(a) \equiv \frac{1}{3\bar{\rho}(a)} \langle 	\delta^{ij} \tau^{\textrm{eff}}_{ij}(a,\vec{x}) \rangle = \frac{1}{3} \left[2\kappa(a) + u(a)\right] \,,
\end{equation} 
where we combined Eqs.~(\ref{eq:effstress}),(\ref{eq:kappa}) and (\ref{eq:Omega}). Note that with this definition the cosmic energy Eq.~(\ref{eq:cosmic_energy}) can be written in a more familiar form:
\begin{equation}
	\bar{\rho}_{\textrm{total}}' + 3\mathcal{H} 	\bar{\rho}_{\textrm{total}}(1+\omega) = 0 \,,
\end{equation}
where $\bar{\rho}_{\textrm{total}}=\bar{\rho}(1+\kappa +u)$ is the total background energy density, including contributions from the backreaction of short distance fluctuations, and we used $\bar{\rho} \propto a^{-3}$. 

From Eq.~(\ref{eq:eos}) we also see that virialized scales, for which $2\kappa + u=0$, exactly decouple and do not contribute to the averaged effective stress \cite{Peebles:2009hw}. This is much stronger than the statement that contributions from virialized structures are parametrically suppressed, which perhaps would be more natural from an EFT point of view. Instead, the decoupling of virialized scales should be thought of as a  non-renormalization theorem \cite{Baumann:2010tm}.

\begin{figure}
	\centering
	\includegraphics[width=0.75\textwidth]{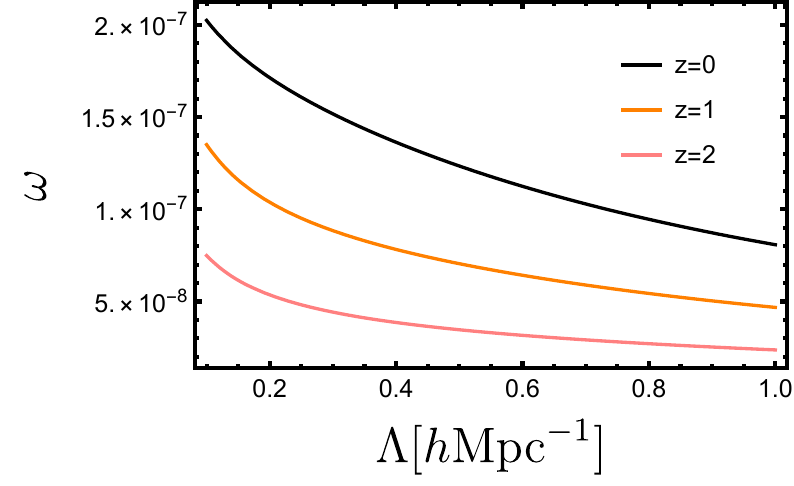}
	\caption{Equation of state as a function of the cutoff scale $\Lambda$ in our fiducial cosmology.}
	\label{fig:eos}
\end{figure}

With the ingredients developed in Sec.\ref{sec:eftoflss}, particularly Eqs.~(\ref{eq:binding_calculated}) and (\ref{eq:kit_int}), we can now easily compute the equation of state. This is plotted in Fig.~\ref{fig:eos}, as a function of the cutoff scale $\Lambda$, in our fiducial $\Lambda$CDM$_\textrm{DarkSky}$ cosmology. 

Note that the value of $\omega \approx 2 \times 10^{-7}$, at $R=\Lambda^{-1}=10h^{-1}\textrm{Mpc}$ and $z=0$, translates to a velocity of $\sqrt{2\omega} \approx 190 \textrm{km/s}$, which is about a factor of two smaller than the value of $\sigma_{\textrm{dis}} \approx 450 \textrm{km/s}$ observed in Fig.~\ref{fig:vel_dis} at the same scale and redshift. This is due to (aside from the small difference in cosmologies) a partial cancellation between kinetic and potential contributions to the equation of state, which is exact for virialized scales.

Let us emphasize that our simple calculation for the equation of state makes no approximations, and only assumes accurate knowledge of the fully nonlinear matter power spectrum as a function of redshift \footnote{To compute the equation of state at $z=0$ one only needs to know the matter power spectrum out to $z\lesssim 2$. This is because the time integral in Eq.~(\ref{eq:kit_int}) is dominated by its contributions from $z \ll 1$.}. This is possible due to the fact that at this level all one needs to compute is the averaged effective stress tensor, which reduces to its trace due to statistical isotropy and its two contributions, i.e. kinetic and potential, are actually tied together by the cosmic energy Eq.~(\ref{eq:cosmic_energy}). The situation is more complicated for the effective sound speed (the only 1-loop EFTofLSS free parameter), since in that case what matters is how the averaged stress responds to the presence of a long-wavelength mode. However, we will now see that it is still possible to meaningfully estimate the effective sound speed based on separate universe techniques.

\section{Effective sound speed}
\label{sec:eff_cs}

In Sec.\ref{sec:eftoflss} we wrote down Eqs.~(\ref{eq:effstress})-(\ref{eq:potential}) for the effective stress tensor,  $\tau^{\textrm{eff}}_{ij}(a,\vec{x})$, in the EFTofLSS. In principle this fully specifies the effective sound speed in terms of short-wavelength fluctuations, but in practice one traditionally follows a bottom-up approach that consists in writing down an expansion including all operators made of long-wavelength fluctuations that are consistent with the symmetries \cite{Kehagias:2013yd, DAmico:2021rdb, Marinucci:2024add}, and with arbitrary time-dependent coefficients. This reads \footnote{This expansion is local in time, which is sufficient at 1-loop order. See \cite{Ansari:2024efj} for the more general case of time non-locality, and references therein for recent discussions on the role of time non-locality in cosmological perturbation theory \cite{Donath:2023sav}.}
\begin{equation}
	\label{eq:eft_exp}
	\frac{\tau^{\textrm{eff}}_{ij}}{\bar{\rho}} = \omega \delta_{ij} + \left(c_{s}^2 \delta_{l} - \frac{c_{bv}^2}{aH} \vec{\nabla} \cdot \vec{v}_{l} \right)\delta_{ij} -\frac{3}{4} \frac{c_{sv}^2}{aH} \left(\partial_{j} v_{l i} + \partial_{i} v_{l j} -\frac{2}{3} \delta_{ij} \vec{\nabla} \cdot \vec{v}_{l} \right) + \Delta \tau_{ij} + \dots  \,,
\end{equation}
where $\omega$ is the equation of state calculated in Sec.\ref{sec:eos}, $c_{s}^2(a)$ is the speed of sound, $c_{bv}^2(a)$ and $c_{sv}^2(a)$ are the coefficients of bulk and shear viscosity respectively, $ \Delta \tau_{ij}$ represents stochastic terms and there are additional contributions of higher order in long-wavelength fields and derivatives thereof. 

The relevant quantity, which appears in the Euler Eq.~(\ref{eq:euler2}) (the divergence of that equation to more precise), is
\begin{equation}
	\label{eq:eft_exp_source}
	\partial^{i}\left(\frac{1}{\rho_{l}} \partial^{j} \tau^{\textrm{eff}}_{ij} \right) = \frac{1}{\bar{\rho}} \partial^{i} \partial^{j} \tau^{\textrm{eff}}_{ij} + \dots = c_{s}^2 \nabla^2 \delta_{l} - \frac{c_{bv}^2+c_{sv}^2}{aH} \nabla^2(\vec{\nabla} \cdot \vec{v}_{l}) = c_{\textrm{eff}}^{2} \nabla^2 \delta_{l}  \,,
\end{equation}
where
\begin{equation}
	\label{eq:eff_cs}
	c_{\textrm{eff}}^2(a) = c_{s}^2(a) +f(a)\left[c_{bv}^2(a)+c_{sv}^2(a)\right] \,,
\end{equation}
is the effective sound speed. In the second equality in Eq.~(\ref{eq:eft_exp_source}) we dropped higher order and stochastic terms, and used Eq.~(\ref{eq:eft_exp}), and finally the third equality holds at one-loop level in perturbation theory since in that case we can apply the linear theory relation $\vec{\nabla} \cdot \vec{v}_{l}/aH = -f\delta_{l}$, with $f(a)=d\log D_{\textrm{L}}/d\log a$ the linear growth rate. 

The effective sound speed is then to be determined by matching the EFT perturbative power spectrum to the fully nonlinear power spectrum from simulations or observations. However, we will now show that the effective sound speed can be estimated directly from Eqs.~(\ref{eq:effstress})-(\ref{eq:potential}) using tools developed to compute the equation of state in Sec.\ref{sec:eos}, and in combination with separate universe methods. Let us also mention that the sound speed has been computed exactly from Eqs.~(\ref{eq:effstress})-(\ref{eq:potential}) via cross correlations using N-body simulations through a much more complicated procedure (see Section 2.4 in \cite{Carrasco:2012cv}).

Now it no longer suffices to consider the average effective stress tensor since the quantity of interest, the left-hand side of Eq.~(\ref{eq:eft_exp_source}), is proportional to derivatives of that tensor (i.e., these are tidal effects). To proceed, we exploit the separation of scales between long-wavelength and short-wavelength fluctuations to consider a local average of the effective stress tensor in the presence of a long-wavelength fluctuation $\delta_{l}(a,\vec{x})$, $\langle \tau^{\textrm{eff}}_{ij} \rangle_{\delta_{l}(a,\vec{x})}$. This can be estimated using standard separate universe techniques that take the limit of an infinitely long-wavelength homogeneous fluctuation $\delta_{l}(a)$, which can be absorbed in the background expansion \cite{Wagner:2014aka, Baldauf:2015vio}. This implicitly makes the assumption of a large separation of scales, with $k_{s} > \Lambda \gg k_{l}$. This is equivalent to dropping higher derivative corrections, which are not the subject of our investigation in any case.

The averaged stress tensor will only have a trace component in the separate universe due to statistical isotropy (i.e. $\delta_{l}$ is spherically symmetric). We will have more to say about this at the end of this section, but for now we simply consider the trace's contribution to the effective stress
\begin{equation}
	\label{eq:trace}
	\langle \tau^{\textrm{eff}}_{ij} \rangle_{\textrm{SU}} = \frac{1}{3} \delta_{ij} \langle \tau^{\textrm{eff}} \rangle_{SU} \,.
\end{equation}
where we introduce the notation $\tau^{\textrm{eff}} \equiv (\tau^{\textrm{eff}})^{k}_{\ k}$, and quantities evaluated on the separate universe will carry a $\textrm{SU}$ subscript, such as $\langle \tau^{\textrm{eff}} \rangle_{SU}$. The first step to compute this is to expand in the long-wavelength fluctuation
\begin{equation}
	\label{eq:taylorexp}
	\langle \tau^{\textrm{eff}} \rangle_{\textrm{SU}} = \langle \tau^{\textrm{eff}} \rangle + \frac{\partial 	\langle \tau^{\textrm{eff}} \rangle_{\textrm{SU}}}{\partial \delta_{l}}\Big|_{\delta_{l}=0} \delta_{l} + \mathcal{O}(\delta_{l}^2) \,,
\end{equation}
It will prove useful to define the equation of state in the presence of a homogeneous fluctuation, generalizing Eq.~(\ref{eq:eos}). That is,
\begin{equation}
	\label{eq:eoshf}
	\omega_{\textrm{SU}} = \frac{\langle \tau^{\textrm{eff}} \rangle_{\textrm{SU}}}{3\bar{\rho}_{\textrm{SU}}} \,.
\end{equation}

The background evolution in the separate universe differs from the original cosmology. We review some results that will be needed shortly, and refer the reader to \cite{Wagner:2014aka, Li:2014sga} for additional details. The  matter density in the separate universe is given by $\bar{\rho}_{\textrm{SU}}=\bar{\rho}(1+\delta_{l})$. From $\bar{\rho}_{\textrm{SU}} \propto a_{\textrm{SU}}^{-3}$ with $a_{\textrm{SU}}$ the scale factor in the separate universe, and the Friedmann Eq.~(\ref{eq:fried}) evaluated at the present time, it follows that
\begin{equation}
	\label{eq:separateuni}
	\frac{\Omega_{\textrm{m,0}}H_{0}^2}{a^3} (1+\delta_{l}) = 	\frac{\Omega_{\textrm{m,0}}^{\textrm{SU}}(H_{0}^{\textrm{SU}})^2}{a_{\textrm{SU}}^3} \,.
\end{equation}
If at sufficiently early times we switch off the homogeneous fluctuation, $\delta_{l} \to 0$, while also demanding that $a_{\textrm{SU}} \to a$ in the same limit, it then follows that $\Omega_{\textrm{m,0}}H_{0}^2 = \Omega_{\textrm{m,0}}^{\textrm{SU}}(H_{0}^{\textrm{SU}})^2$. This implies
\begin{equation}
	\label{eq:susf}
	a_{\textrm{SU}} = a(1+\delta_{l})^{-\frac{1}{3}} \implies \frac{da_{\textrm{SU}}}{d\delta_{l}}\Big|_{\delta_{l}=0} = -\frac{1}{3} a \,.
\end{equation} 

These are all the results we need, and we are ready to substitute Eq.~(\ref{eq:eoshf}) into Eq.~(\ref{eq:taylorexp}) to obtain
\begin{equation}
	\label{eq:taylorexpagain}
	\langle \tau^{\textrm{eff}} \rangle_{\textrm{SU}} = 3\bar{\rho} \left[\omega + \left( \omega + \frac{d\omega_{\textrm{SU}}}{d\delta_{l}}\Big|_{\delta_{l}=0}\right) \delta_{l}  + \mathcal{O}(\delta_{l}^2) \right] \,,
\end{equation}
where $\omega$ is the equation of state in the original cosmology, and we used the relation $\bar{\rho}_{\textrm{SU}}=\bar{\rho}(1+\delta_{l})$. Combining Eqs.~(\ref{eq:eft_exp_source}), (\ref{eq:trace}) and (\ref{eq:taylorexpagain}) finally yields
\begin{equation}
	\label{eq:effcs}
	c_{\textrm{eff}}^{2} =  \omega + \frac{d\omega_{\textrm{SU}}}{d\delta_{l}}\Big|_{\delta_{l}=0} \,.
\end{equation}
We can now use the results of Sec.\ref{sec:eos} for the equation of state to estimate the effective sound speed. From Eq.~(\ref{eq:eos}),
\begin{equation}
	\label{eq:eoslm}
	\frac{d\omega_{\textrm{SU}}}{d\delta_{l}}\Big|_{\delta_{l}=0} = \frac{1}{3} \left(2\frac{d\kappa_{\textrm{SU}}}{d\delta_{l}}\Big|_{\delta_{l}=0} + \frac{du_{\textrm{SU}}}{d\delta_{l}}\Big|_{\delta_{l}=0} \right) \,.
\end{equation}
Taking a derivative of Eqs.~(\ref{eq:binding_calculated}) and (\ref{eq:kit_int}) with respect to $\delta_{l}$ gives \footnote{The combination $\Omega_{\textrm{m},0}H_{0}^2$ is independent of the long mode as we showed above, but the scale factor is not, and we need to use Eq.~(\ref{eq:susf}) when taking derivatives with respect to the long mode.}, 
\begin{equation}
	\label{eq:pot_int_res}
	\frac{du_{\textrm{SU}}}{d\delta_{l}}\Big|_{\delta_{l}=0} = -\frac{3}{4} \Omega_{\textrm{m},0} H_{0}^2 \frac{1}{a} \int \frac{dk}{2\pi^2} P(a,k) \left[\frac{1}{3} + R(a,k) \right] \left[1-W_{\Lambda}(k) \right]^2 \,,
\end{equation}
and
\begin{equation}
	\label{eq:kit_int_res}
	\frac{d\kappa_{\textrm{SU}}}{d\delta_{l}}\Big|_{\delta_{l}=0} = -\frac{1}{3} u  -\frac{du_{\textrm{SU}}}{d\delta_{l}}\Big|_{\delta_{l}=0} +a^{-2} \int_{0}^{a} da' a' \left[\frac{du_{\textrm{SU}}}{d\delta_{l}}\Big|_{\delta_{l}=0} + \frac{2}{3} u(a') \right]\,,
\end{equation}
respectively, where
\begin{equation}
	\label{eq:response}
	R(a,k) = \frac{d\log P_{\textrm{SU}}(a,k)}{d\delta_{l}}\Big|_{\delta_{l}=0} \,,
\end{equation}
is the response function. This quantity is directly related to the angle-averaged squeezed limit bispectrum \cite{Kehagias:2013paa, Ben-Dayan:2014hsa, sherwin2012shift}, and was extracted from separate universe simulations in \cite{Li:2014sga, Wagner:2015gva, Baldauf:2015vio}. Here we use the semi-analytic halo model (HM) \cite{Cooray:2002dia} to calculate the response, with explicit formulas provided in Appendix \ref{sec:app3}.  Despite the limitations of the halo model, its predictions for the response function are in excellent agreement with the simulation results (see Fig.4 in \cite{Wagner:2015gva} for a comparison at redshifts $z=0$ and $z=2$).  

We now have all the ingredients needed to compute the estimated effective sound speed from separate universe techniques using  Eq.~(\ref{eq:effcs}), and we do so for two distinct cosmologies, matching the different choices made in a couple of previous independent measurements of the effective sound speed from N-body simulations. The first cosmology, here denoted by $\Lambda CDM_{\textrm{Consuelo}}$, is defined by: $\Omega_{\textrm{m}} =0.25$, $\Omega_{\Lambda}=0.75$, $h=0.7$, $\sigma_{8}=0.8$ and $n_{s}=1$, matching the choice made in \cite{Carrasco:2012cv} which uses the Consuelo simulations \cite{Behroozi_2012}. The second one is our fiducial cosmology, $\Lambda CDM_{\textrm{DarkSky}}$: $\Omega_{\textrm{m}} =0.295$, $\Omega_{\Lambda}=0.705$, $h=0.688$, $\sigma_{8}=0.835$ and $n_{s}=0.9676$, matching the choice made in \cite{Foreman:2015lca} which uses the Dark Sky simulations \cite{Skillman:2014qca}.

\begin{figure}
	\centering
	\includegraphics[width=1\textwidth]{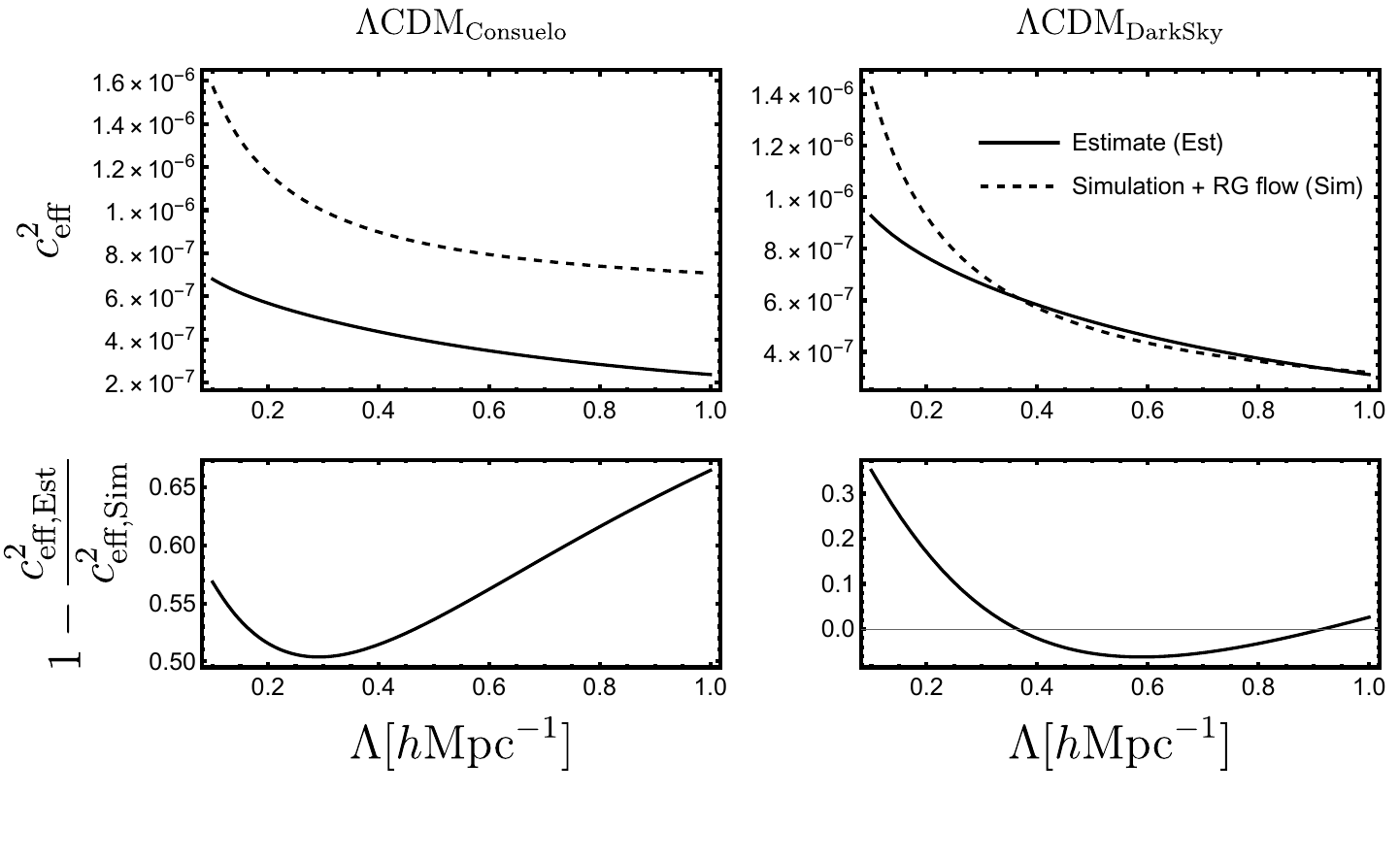}
	\caption{Effective sound speed as a function of the cutoff scale $\Lambda$, at redshift $z=0$. Solid lines correspond to the estimate from separate universe techniques, and dashed lines are obtained from matching to simulations with a cutoff dependence fixed by the RG flow. Lower plots show the relative difference between estimated and simulation results. Left and right panels compare our estimate to the simulation results of \cite{Carrasco:2012cv} and \cite{Foreman:2015lca}. There is better agreement (with differences at the few tens of percent level) between the estimated sound speed and the higher resolution simulation result of \cite{Foreman:2015lca} (right panel), when compared to the lower resolution simulation result of \cite{Carrasco:2012cv} (left panel). Additional comments on the discrepancy between estimated and lower resolution simulation results, as seen on the left panel, can be found in the main text. }
	\label{fig:ceff}
\end{figure}

In Fig.~\ref{fig:ceff} we plot the effective sound speed as a function of the cutoff scale $\Lambda$, at redshift $z=0$. Solid lines show the separate universe estimate and dashed lines are obtained from matching to simulations in the renormalized limit of $\Lambda \to \infty$, with a cutoff dependence fixed by the Renormalization Group (RG) flow (see Appendix \ref{sec:app2} for details). The left and right panels compare our estimate to the simulation results of \cite{Carrasco:2012cv} and \cite{Foreman:2015lca} respectively, \footnote{Many other works have measured the sound speed counterterm in different scenarios, including a scaling universe \cite{Carrasco:2013sva}, simplified initial conditions in Einstein-de Sitter \cite{Karandikar:2023ozp}, Lagrangian perturbation theory \cite{Baldauf:2015tla}, in 1+1 dimensions \cite{McQuinn:2015tva} and with an IR resummed theoretical template \cite{Ivanov:2018lcg, Senatore:2014via}. Also see \cite{Angulo:2014tfa, Foreman:2015uva} and some works on the EFT at the field level, e.g. \cite{Kostic:2022vok, Tucci:2023bag, Schmidt:2020ovm, Schmittfull:2018yuk, Nguyen:2024yth}.} with details on these comparisons regarding different conventions in the literature explained in Appendix \ref{sec:app2}. 

It is important to note that \cite{Carrasco:2012cv} was one of the very first efforts to measure the effective sound speed, using low resolution simulations and matching at a high renormalized scale of $k_{\textrm{ren}}=0.16h\textrm{Mpc}^{-1}$. On the other hand, the follow-up work of \cite{Foreman:2015lca} used higher resolution simulations and aimed for precision, managing to match at much lower scales of $k_{\textrm{ren}} \lesssim 0.05h\textrm{Mpc}^{-1}$. One of the conclusions reached in \cite{Baldauf:2015aha} was precisely that the effective sound speed can be overestimated by as much as a factor of two when matching at a high scale of $k_{\textrm{ren}} \approx 0.2h\textrm{Mpc}^{-1}$. This is apparent in the left plot of Fig.~\ref{fig:ceff}, where the estimated effective sound speed is about a factor of two lower than the low resolution simulation result, which is likely overestimated by the aforementioned reasons. On the other hand, as shown in the right panel in Fig.~\ref{fig:ceff}, there is a much better agreement between estimated and higher resolution simulation effective sound speeds.

When interpreting Fig.~\ref{fig:ceff} it is important to keep in mind, as we will expand upon at the end of this section, that our framework is limited in that it does not account for the traceless part of the effective stress tensor. We expect its contribution to be comparable in size to the one from the trace, and hence the estimates provided to be correct up to a factor of two. In particular, we do not expect the contribution from the trace to separately track the RG flow.

\begin{figure}
	\centering
	\includegraphics[width=0.75\textwidth]{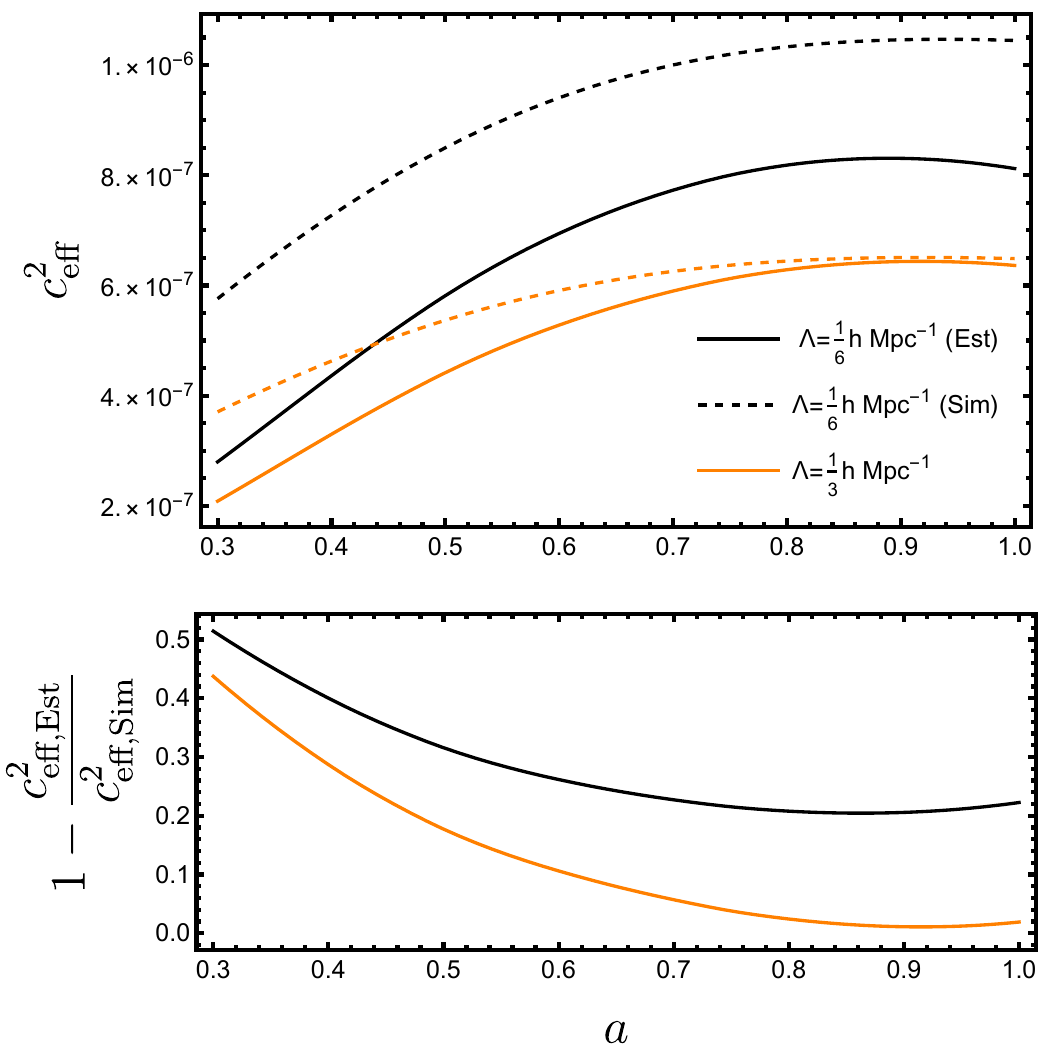}
	\caption{Effective sound speed as a function of the scale factor in our fiducial cosmology. Solid lines show the separate universe estimate while dashed lines correspond to the simulation results of \cite{Foreman:2015lca} when augmented with the RG flow plus a scaling universe ansatz for the time dependence of the renormalized counterterm (see Appendix \ref{sec:app2} for more details). Black and orange lines show results for $\Lambda=(1/6) h \textrm{Mpc}^{-1}$ and $\Lambda=(1/3) h \textrm{Mpc}^{-1}$, respectively. The time dependence of the estimated sound speed is approximately consistent with the assumption of a scaling universe, which was verified to accurately capture the time-dependence from full simulations. }
	\label{fig:ceff_time}
\end{figure}

In Fig.~\ref{fig:ceff_time} we plot the estimated and simulation effective sound speeds as a function of the scale factor in our fiducial cosmology, for two different values of the cutoff, $\Lambda=(1/6) h \textrm{Mpc}^{-1}$ and $\Lambda=(1/3) h \textrm{Mpc}^{-1}$. Let us mention that the interpretation of simulation results (here from \cite{Foreman:2015lca}) require an assumption about the time dependence of the renormalized effective sound speed counterterm, which we take to be that expected from a scaling universe \cite{Pajer:2013jj} (and has been found to be consistent with the time dependence from N-body simulations \cite{Baldauf:2014qfa, Foreman:2015uva}). We elaborate on this in Appendix \ref{sec:app2}. Fig.~\ref{fig:ceff_time} shows that the time dependence of the estimated sound speed is approximately consistent with the assumption of a scaling universe.

Let us stop for a moment to address a potential source of confusion related to the comparison of our semi-analytic estimate to simulation results. A naive direct application of our formalism in the limit $\Lambda \to \infty$ produces a vanishing effective sound speed, since in that limit there are no short modes contributing to the counterterm, while one does find a nonzero value for the renormalized counterterm from simulations in the same limit. The reason for this is that in our approach it does not make sense to directly push the cutoff far beyond the scale of nonlinearities, since then one would lose the perturbative control over the long-wavelength modes. On the other hand, theoretical control over all scales can be achieved with simulations, so a similar issue does not arise in this case \footnote{In reality simulations have a finite resolution and make approximations such as force softening, so theoretical control can only be achieved up to some cutoff scale $\Lambda_{\textrm{sim}}$. This is of no practical importance as long as the observables of interest are well converged with respect to this cutoff.}.

More concretely, in the EFTofLSS the sound speed counterterm accomplishes two things at the same time. It accounts for the backreactions of short-distance fluctuations into the long modes, but it also absorbs the UV dependence of perturbation theory loop integrals. By working with a finite cutoff, which is below the scale of nonlinearities, one can ensure that the latter contributions to the counterterm are effectively vanishing since only perturbative modes are running in the loops. Then we only need to account for the former, physical contribution due to the coupling of short and long wavelength modes which is exactly the quantity we aim to estimate in this work. Afterwards, the unphysical contribution from the spurious UV behavior of loop integrals can be encapsulated by RG flowing towards $\Lambda= \infty$. 

Note that due to the arguments made in the previous paragraph, in Fig.~\ref{fig:ceff} we choose to compare the estimated and simulation sound speeds at a finite cutoff. An alternative approach would be to RG flow the estimated sound speed towards $\Lambda = \infty$, to compare with the renormalized counterterms directly extracted from simulations. However, since the estimates do not exactly follow the RG flow as shown in Fig.~\ref{fig:ceff}, this procedure would result in an estimated renormalized sound speed that depends on a spurious reference scale $\Lambda_{\textrm{ini}}$ (the inital scale one is running from, towards $\Lambda \to \infty$). For example, in the Dark Sky cosmology at $z=0$, the choices of $\Lambda_{\textrm{ini},1}=0.3h\textrm{Mpc}^{-1}$ and $\Lambda_{\textrm{ini},2}=0.5h\textrm{Mpc}^{-1}$ produce $c_{\textrm{eff},1}^{2}(\Lambda=\infty) \approx 1.4 \times 10^{-7}$ and $c_{\textrm{eff},2}^{2}(\Lambda=\infty) \approx 2 \times 10^{-7}$. We avoid this ambiguity by RG flowing the simulation results to finite $\Lambda$ to compare with our estimates.
 
\begin{figure}
	\centering
	\includegraphics[width=0.75\textwidth]{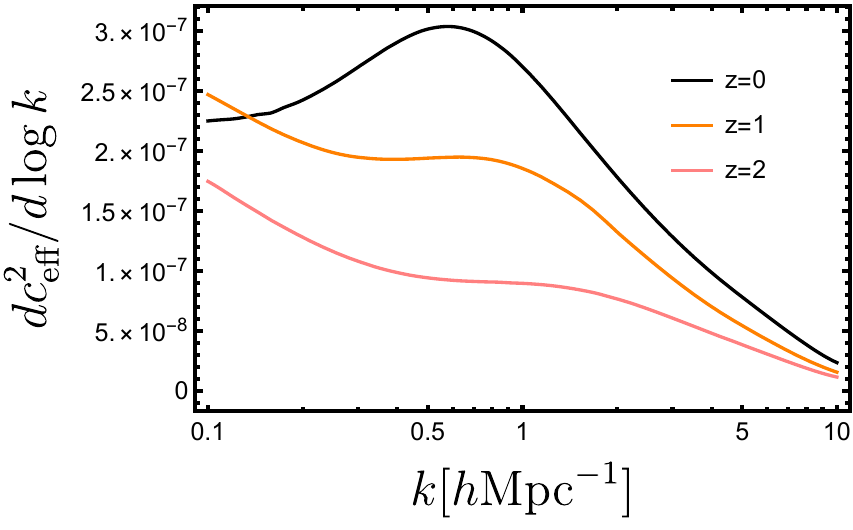}
	\caption{Derivative of the effective sound speed with respect to wavenumber, as defined in Eq.~(\ref{eq:cs_diff}). The features in the plot directly trace the shape of the derivative of the power spectrum with respect to a long mode evaluated at $\delta_{l}=0$, and show that scales as small as a few $\sim h\textrm{Mpc}^{-1}$ have a sizable contribution to the effective sound speed. Additionally, the peaks at $k\sim (0.5-1)h\textrm{Mpc}^{-1}$ suggest that cosmic sheets and filaments are the structures which contribute the most to the effective sound speed.}
	\label{fig:cs_diff}
\end{figure}

We find it remarkable that a semi-analytic calculation can reproduce the effective sound speed extracted from simulations so well. Let us emphasize that the short scale gravitational binding energy in the separate universe is the basic ingredient that goes into estimating the effective sound speed, or exactly computing the equation of state, and hence these calculations facilitate the interpretability of the EFT counterterm. For example, let us introduce the quantity $dc_{\textrm{eff}}^{2}/d\log k$ according to
\begin{equation}
\label{eq:cs_diff}
	c_{\textrm{eff}}^2 = \int_{k>\Lambda} d\log k \frac{dc_{\textrm{eff}}^{2}}{d\log k} \,,
\end{equation}
which carries information about what scales contribute the most to the effective sound speed, and can be easily extracted by differentiating the latter with respect to the cutoff. This is plotted in Fig.~\ref{fig:cs_diff}. The features in the plot directly trace the shape of the power spectrum times the response function, i.e. the derivative of the power spectrum with respect to a long mode evaluated at $\delta_{l}=0$. The curves for $dc_{\textrm{eff}}^{2}/d\log k$ show a peak at typical scales of sheets and filaments, $k \sim (0.5-1)h\textrm{Mpc}^{-1}$. This suggests that these are the cosmic structures which contribute the most to the effective sound speed. While we expect the contributions from the traceless part of the stress tensor to make quantitative changes to the shape of the curves in Fig.~\ref{fig:cs_diff} (for example, it may somewhat shift the peak locations), we expect the overall qualitative features to remain the same.

\begin{figure}
	\centering
	\includegraphics[width=0.75\textwidth]{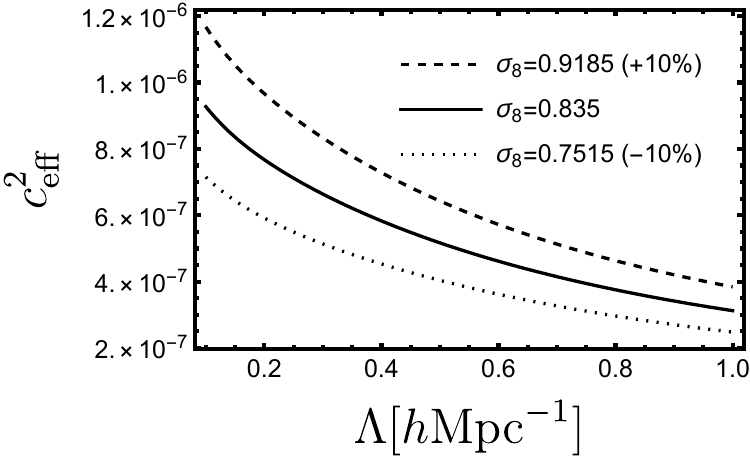}
	\caption{Effective sound speed as a function of the cutoff scale for varying $\sigma_{8}$. Here $z=0$ and we consider $10\%$ variations on  $\sigma_{8}$ away from its value in our fiducial cosmology. This corresponds to an approximate scaling of $c_{\textrm{eff}}^2 \sim \sigma_{8}^{2.2-2.5}$.}
	\label{fig:cs_sigma8}
\end{figure}

\begin{figure}
	\centering
	\includegraphics[width=0.75\textwidth]{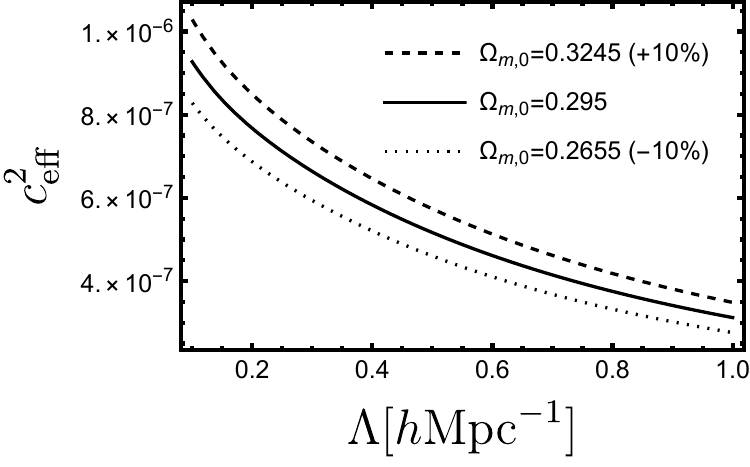}
	\caption{Effective sound speed as a function of the cutoff scale for varying $\Omega_{\textrm{m},0}$. Here $z=0$ and we consider $10\%$ variations on $\Omega_{\textrm{m},0}$ away from its value in our fiducial cosmology. This corresponds to an approximate scaling of $c_{\textrm{eff}}^2 \sim \Omega_{\textrm{m},0}^{1.1-1.2}$.}
	\label{fig:cs_omegam}
\end{figure}

Additionally, the EFT counterterm carries information about the short scale gravitational dynamics, and is hence sensitive to the cosmology. This can be easily investigated within our semi-analytic framework. In Fig.~\ref{fig:cs_sigma8} we show the response of the effective sound speed to varying $\sigma_{8}$, which corresponds roughly to a scaling of $c_{\textrm{eff}}^2 \sim \sigma_{8}^{2.2-2.5}$. Fig.~\ref{fig:cs_omegam} shows the response of the effective sound speed to varying $\Omega_{\textrm{m,0}}$, corresponding roughly to a scaling of $c_{\textrm{eff}}^2 \sim \Omega_{\textrm{m,0}}^{1.1-1.2}$.

A strong dependence on $\sigma_{8}$ was previously pointed out in \cite{Foreman:2015uva} based on simulations, where they report a $c_{\textrm{ctr}} \sim \sigma_{8}^{3.5}$ scaling for the counterterm (see footnote 13 in that reference). However, their comparison is not done while holding the other cosmological parameters fixed (for example, the cosmology with higher $\sigma_{8}$ also has a value of $\Omega_{\textrm{m,0}}$ which is $\sim 5\%$ larger). Additionally, they extract the counterterm in the renormalized limit ($\Lambda \to \infty$) while we compute the sound speed at a finite cutoff, so a direct comparison is not appropriate.

We finish this section with a discussion on the main limitations of the separate universe approach pursued here. It cannot account for higher derivative corrections due to the underlying assumption of a large separation of scales, i.e.,  $k_{s} > \Lambda \gg k_{l}$. It does not account for stochastic terms either since it is only upon averaging that we can connect the kinetic and potential terms of the effective stress tensor, via the cosmic energy equation. These are not relevant limitations if one is only interested in the effective sound speed.

Most importantly, our approach misses contributions from the traceless part of the effective stress tensor, see Eq.~(\ref{eq:trace}). The local averaged traceless stress is nonzero precisely because at each point in space the preferred vector $\vec{\nabla} \delta_{l}$ locally breaks isotropy. In fact, previous measurements from simulations \cite{Carrasco:2012cv} and analytic estimates from perturbation theory considerations (see Appendix D in \cite{Baumann:2010tm}) indicate the contribution from the traceless part of the effective stress tensor and from its trace to be of the same order of magnitude. We can then expect errors that may be as large as order unity, which is reflected in Figs.~\ref{fig:ceff} and \ref{fig:ceff_time}. This issue can potentially be addressed with anisotropic separate universe simulations, which account for the presence of a large-scale tidal field \cite{Masaki:2020drx, Stucker:2020fhk, Li:2017qgh, Akitsu:2020fpg}, and may allow for an exact calculation of the effective sound speed using separate universe techniques (as first suggested in \cite{Garny:2015oya}).

\section{Conclusion}
\label{sec:concl}

Effective field theory methods for large-scale structure have proven themselves to be extremely powerful in pushing the regime of applicability of cosmological perturbation theory to smaller scales, and are now a central piece of the theory modeling involved in the analysis of real data. Even in its simplest form \footnote{In the context of modeling the power spectrum of the matter field, in real space, at 1-loop in perturbation theory. This set-up is assumed in the discussion that follows.}, the EFTofLSS comes at the cost of adding a new nuisance parameter to the theory, the effective sound speed, which can be extracted from cosmological simulations by matching EFT predictions to the fully nonlinear power spectrum.

While very effective, this procedure does not provide a clear physical interpretation for the counterterm in that it does not shed light on the microphysics that is ultimately responsible for the effective sound speed.  This is to be contrasted, for example, with Chiral perturbation theory in QCD \cite{Gasser:1983yg} which has the pion decay constant as a free parameter (a quantity that has a physical interpretation). In this work we fill that gap by providing a semi-analytic estimate of the effective sound speed from a simplified model of the short scale fluctuations based on separate universe techniques. Our method reproduces the results from simulations with errors at the few tens of percent level at redshifts $z\lesssim 2$, and across a range of cutoffs $ 0.1h\textrm{Mpc}^{-1} \leq \Lambda \leq 1h\textrm{Mpc}^{-1}$. This is summarized in Figs.~\ref{fig:ceff} and \ref{fig:ceff_time}. 

In Sec.\ref{sec:eftoflss} we review the EFT approach to large scale structure and emphasize the power of the Layzer-Irvine Eq.~(\ref{eq:cosmic_energy}), which enables a calculation of the average velocity dispersion, smoothed on a given scale, assuming knowledge of the fully nonlinear power spectrum, see Fig.~\ref{fig:vel_dis}. We build upon this to first compute the equation of state in Sec.\ref{sec:eos}, as summarized in Fig.~\ref{fig:eos}. Finally, in Sec.\ref{sec:eff_cs} we use the separate universe approach to semi-analytically estimate the effective sound speed. The basic ingredient needed in that calculation is the short-scale gravitational binding energy in a separate universe, which provides a physical interpretation for the counterterm. The effective sound speed carries information about the nonlinear gravitational dynamics of short-scale fluctuations, revealed for example by its dependence on cosmological parameters, see Figs.~\ref{fig:cs_sigma8} and \ref{fig:cs_omegam}. This is underscored in our approach which is ultimately based on integrals of the fully nonlinear power spectrum with contributions from very small scales, as illustrated in Fig.~\ref{fig:mps} and \ref{fig:cs_diff}. 

In a broader note, methods that shed light on and/or constrain nuisance parameters in our EFT based perturbartive models of large scale structure can be helpful in providing well informed theoretical priors for such parameters. This improves the extraction of cosmological parameters \cite{Ivanov:2024xgb, Zhang:2024thl, Ivanov:2024hgq}, and helps to alleviate prior volume effects in Bayesian analyses \cite{Ivanov:2019pdj, Philcox:2021kcw, Ivanov:2023qzb, Holm:2023laa, Simon:2022lde, Maus:2024sbb, Carrilho:2022mon, Donald-McCann:2023kpx}.

We have some ideas for future directions of investigation. It would be interesting to further elucidate the physical interpretation of the equation of state and effective sound speed. A field level analysis could help to provide a visual picture of what type of nonlinear structures contribute the most to these two quantities, such as sheets or filaments \footnote{This type of analysis may however be hindered by the fact that differentiating these kinds of structures is still an open area of research (see, e.g., \cite{Feldbrugge:2024wcm, Dhawalikar:2024ymt, palomino2024cosmic,Feldbrugge:2024wsb, Sarkar:2024jpg, Khoshtinat:2024tfr, Ilc:2024rrm} for a few recent papers on the subject).}, and directly reveal the decoupling of fully virialized structures in an instructive way. Additionally, it would be worthwhile to test the limits of our approach in a thorough comparison against different methods to extract the sound speed from simulations in different cosmologies, potentially even beyond $\Lambda$CDM. Such investigations hold the potential to reveal interesting aspects of the short scale gravitational dynamics underlying the EFTofLSS.

\acknowledgments
This work is supported by the Department of Energy grants DE-SC0023183 and DE-SC0011637 and the Dr. Ann Nelson Endowed Professorship. MM is supported by NSF award AST-2007012. M.L. and C.N are grateful for the hospitality of Perimeter Institute where part of this work was carried out.
Research at Perimeter Institute is supported in part by the Government of Canada through the
Department of Innovation, Science and Economic Development Canada and by the Province of
Ontario through the Ministry of Economic Development, Job Creation and Trade. This research
was also supported in part by the Simons Foundation through the Simons Foundation Emmy
Noether Fellows Program at Perimeter Institute.  We thank the MilleniumTNG collaboration for giving us permission to use their simulations data. DJ thanks Eiichiro Komatsu and Volker Springel for providing helpful comments. CN thanks Charuhas Shiveshwarkar for many helpful discussions, and Mikhail Ivanov for helpful comments on a draft. All the numerical calculations and plots in this paper were made with Mathematica \cite{Mathematica}.

\appendix 
 
\section{A derivation of the Layzer-Irvine equation}
\label{sec:app1}

Our goal is to derive the Layzer-Irvine Eq.~(\ref{eq:cosmic_energy}) directly from the Boltzmann Eq.~(\ref{eq:vlasovlong2}), which we repeat here for convenience
\begin{equation}
\label{eq:vlasovapp}
	\frac{\partial f_{l}}{\partial \eta} + \vec{q} \cdot \frac{\partial f_{l}}{\partial \vec{x}} = a^{2}(\eta) \frac{\partial \phi_{l}}{\partial \vec{x}} \cdot \frac{\partial f_{l}}{\partial \vec{q}} + a^2(\eta) \left[\frac{\partial \phi_{s}}{\partial \vec{x}} \cdot \frac{\partial f_{s}}{\partial \vec{q}} \right]_{\Lambda} + O\Big(\frac{k^2}{\Lambda^2}\Big) \,.
\end{equation}
We first introduce the background long-wavelength distribution function, defined by  $\bar{f}_{l}(\eta,q)=\langle f_{l}(\eta,\vec{x},\vec{q}) \rangle$, and its fluctuations $\delta f_{l} = f_{l} - \bar{f}_{l}$. Now take the ensemble average of Eq.~(\ref{eq:vlasovapp}) to obtain
\begin{equation}
\label{eq:vlasovapp-step1}
	\frac{\partial \bar{f}_{l}}{\partial \eta}  = a^2 \left\langle \frac{\partial \phi_{l}}{\partial \vec{x}} \cdot \frac{\partial \delta f_{l}}{\partial \vec{q}} \right\rangle + a^2 \left\langle \frac{\partial \phi_{s}}{\partial \vec{x}} \cdot \frac{\partial f_{s}}{\partial \vec{q}} \right\rangle  + \dots \,,
\end{equation}
where we omit cutoff and time dependences for simplicity of notation, and the ellipsis denote higher derivative corrections.  Each term on the right-hand side of Eq.~(\ref{eq:vlasovapp-step1}) act as a source term, whose contribution to $\bar{f}_{l}$ can be found by integrating it over superconformal time. This leads to a solution of the form
\begin{equation}
\label{eq:vlasovapp-step2}
	\bar{f}_{l}(\eta,q) = \bar{f}_{l,\textrm{pt}}(\eta,q) + \bar{f}_{l,\textrm{ctr}}(\eta,q) + \dots \,,
\end{equation}
where $\bar{f}_{l,\textrm{pt}}(\eta,q)$ and $\bar{f}_{l,\textrm{ctr}}(\eta,q)$ are sourced by the long and short wavelength fluctuations respectively, as given by the first and second terms on the right-hand side of Eq.~(\ref{eq:vlasovapp-step1}). The mixed long-short terms, i.e. higher derivative corrections, vanish in the renormalized limit of $\Lambda \to \infty$ but are otherwise necessary to accurately model observables of interest \cite{Baumann:2010tm, Carrasco:2012cv}. We neglect higher derivative corrections, but note that averages of such terms are always small due to the separation of scales implied by the EFT \footnote{For example, $\langle \phi_{l} \phi_{s} \rangle$ is given by an integral over wavenumber which contains a factor of $W_{\Lambda}(k)[1-W_{\Lambda}(k)]$ in the integrand. This is zero when both $k \ll \Lambda$ and $k \gg \Lambda$. }. 

The first equation we need to solve is then,
\begin{equation}
\label{eq:vlasovapp-step3}
	\frac{\partial \bar{f}_{l,\textrm{pt}}}{\partial \eta} = a^2 \left\langle \frac{\partial \phi_{l}}{\partial \vec{x}} \cdot \frac{\partial \delta f_{l}}{\partial \vec{q}} \right\rangle \,. 
\end{equation}
Since this only involves long-wavelength fluctuations, Eq.~(\ref{eq:vlasovapp-step3}) can be solved perturbatively when coupled to the Boltzmann equation for the distribution function fluctuations as well. A subset of us pursue this in detail elsewhere \cite{Nascimento:2024hle}, and we find that Standard Perturbation Theory (SPT) can be formulated directly at the distribution function level in phase-space, circumventing the need to truncate the Boltzmann hierarchy. Here we only need the following result
\begin{equation}
\label{eq:vlasovapp-step4}
	a^{-5} \int \frac{d^3\vec{q}}{(2\pi)^{3}} q^2 \bar{f}_{l,\textrm{pt}} = \langle \rho_{l} v_{l}^2 \rangle \,,
\end{equation}
which can be understood as follows: SPT does not account for velocity dispersion, hence Eq.~(\ref{eq:vlasovapp-step4}) must hold to ensure that the ensemble average of Eq.~(\ref{eq:kinetic}) vanishes when there are no contributions from short-wavelength fluctuations. 

Moving on to the counterterm contribution to the long-wavelength background distribution function, we need to solve the following equation,
\begin{equation}
\label{eq:vlasovapp-step5}
	\frac{\partial \bar{f}_{l,\textrm{ctr}}}{\partial \eta}  = a^2 \left\langle \frac{\partial \phi_{s}}{\partial \vec{x}} \cdot \frac{\partial f_{s}}{\partial \vec{q}} \right\rangle \,. 
\end{equation}
Now multiply Eq.~(\ref{eq:vlasovapp-step5}) by $q^2$, followed by integrating over momentum $\vec{q}$. The left-hand side of Eq.~(\ref{eq:vlasovapp-step5}) becomes
\begin{equation}
\label{eq:vlasovapp-step6}
	\int \frac{d^3 \vec{q}}{(2\pi)^{3}} q^2 \frac{\partial \bar{f}_{l,\textrm{ctr}}}{\partial \eta} = \frac{d}{d\eta} \int \frac{d^3 \vec{q}}{(2\pi)^{3}} q^2 \bar{f}_{l,\textrm{ctr}} = 2 \frac{d}{d\eta} \left(a^5 \bar{\rho} \kappa \right) \,,
\end{equation}
where the third equality follows from taking the ensemble average of Eq.~(\ref{eq:kinetic}), using Eqs.~(\ref{eq:kappa}), (\ref{eq:vlasovapp-step2}) and (\ref{eq:vlasovapp-step4}). The right-hand side of Eq.~(\ref{eq:vlasovapp-step5}) leads to
\begin{equation}
\label{eq:vlasovapp-step7}
	\int \frac{d^3 \vec{q}}{(2\pi)^{3}} q^2 a^{2} \left\langle \frac{\partial \phi_{s}}{\partial \vec{x}} \cdot \frac{\partial f_{s}}{\partial \vec{q}} \right\rangle = a^2 \left\langle  \frac{\partial \phi_{s}}{\partial \vec{x}} \cdot \int \frac{d^3 \vec{q}}{(2\pi)^{3}} q^2 \frac{\partial f_{s}}{\partial \vec{q}} \right \rangle =  -2a^{6} \left \langle \frac{\partial \phi_{s}}{\partial \vec{x}} \cdot \vec{\Pi}_{s} \right \rangle \,,
\end{equation}
where after integrating by parts we used the short-wavelength analogue of Eq.~(\ref{eq:momentum}). One can now use the short-wavelength analogs of Eqs.~(\ref{eq:poisson}) and (\ref{eq:continuity}) \footnote{All equations which are linear in the distribution function are satisfied separately for both long and short wavelength parts.} to derive the following result, through a series of integration by parts under the average sign (see \footref{footnote2})
\begin{equation}
\label{eq:vlasovapp-step9}
	-2a^{6} \left \langle \frac{\partial \phi_{s}}{\partial \vec{x}} \cdot \vec{\Pi}_{s} \right \rangle = -2a \frac{d}{d\eta} \left(a^4 \bar{\rho} \omega \right) \,,
\end{equation}
where the short scale gravitational binding energy, $v(\eta)$, is defined in Eq.~(\ref{eq:binding}). It is now straightforward to show that combining Eqs.~(\ref{eq:vlasovapp-step6}) and (\ref{eq:vlasovapp-step9}) leads to the desired result:
\begin{equation}
	\label{eq:cosmic_energy_app}
	\frac{d}{d\eta} (\kappa + v) + \mathcal{H} (2\kappa + v) = 0 \,,
\end{equation}
using $\bar{\rho} \propto a^{-3}$ and $\mathcal{H}=d\log a/d\eta$. 

\section{The renormalization group flow}
\label{sec:app2}

In Figs.~\ref{fig:ceff} and \ref{fig:ceff_time} we compare the semi-analytic estimate for the effective sound speed, based on separate universe methods, to the results obtained from matching the EFT to full N-body simulations. In the latter case we read off the renormalized effective sound speed, corresponding to the $\Lambda \to \infty$ limit, directly from the literature \cite{Carrasco:2012cv, Foreman:2015lca, Baldauf:2015aha} and the cutoff dependence is fixed by the RG flow which we now review. 

Our first step will be to use the fluid equations in the EFTofLSS to make the connection between the effective sound speed and the EFT counterterm \footnote{It is often the case in the literature that the EFT counterterm directly defines the effective sound speed. We are following the conventions of \cite{Carrasco:2012cv} where these two quantities are interconnected, but are not the same. This will be made more clear in what follows.}. In terms of $\rho_{l} = \bar{\rho}(1+\delta_{l})$, and $\vec{\Pi}_{l} = \rho_{l} \vec{v}_{l} =  \bar{\rho}(1+\delta_{l})\vec{v}_{l}$, the continuity Eq.~(\ref{eq:continuity}) reads
\begin{equation}
\label{eq:continuityapp}
	\delta_{l}' +a\theta_{l} = -a\vec{\nabla} \cdot (\delta_{l} \vec{v}_{l}) \,,
\end{equation} 
where we used the relation $\bar{\rho} \propto a^{-3}$, and defined $\theta_{l} = \vec{\nabla} \cdot \vec{v}_{l}$. Time, scale and cutoff dependencies are implicit for simplicity of notation. Next we want to take the divergence of the Euler Eq.~(\ref{eq:euler2}), which yields
\begin{equation}
\label{eq:eulerapp}
	\theta_{l}' + \mathcal{H} \theta_{l} + \frac{3}{2} \Omega_{\textrm{m,0}} H_{0}^2 \delta_{l} = -a\partial^{i}\left(v_{l,j} \partial^{j} v_{l,i}\right) -a\partial^{i}\left(\frac{1}{\rho_{l}} \partial^{j} \tau_{ij}^{\textrm{eff}}\right) \,,
\end{equation}
where we used the Poisson Eq.~(\ref{eq:poisson2}). Here we will focus on the contribution to the density contrast coming from the effective stress. Let us first repeat here Eq.~(\ref{eq:eft_exp_source}) defining the effective sound speed:
\begin{equation}
	\label{eq:eftcsapp}
	\partial^{i}\left(\frac{1}{\rho_{l}} \partial^{j} \tau^{\textrm{eff}}_{ij} \right) = c_{\textrm{eff}}^{2} \nabla^2 \delta_{l}  \,.
\end{equation}
The coefficient $c_{\textrm{eff}}^{2}(a)$ is generated by short-wavelength quadratic nonlinearities and hence should be thought of as a second order quantity in perturbation theory to leading order. At one-loop precision we can then substitute $\delta_{l} \to \delta_{l}^{(1)}$ in Eq.~(\ref{eq:eftcsapp}). Keeping that in mind, and dropping all other nonlinear source terms in Eqs.~(\ref{eq:continuityapp}) and (\ref{eq:eulerapp}), we arrive at the following system of equations
\begin{equation}
\label{eq:fluid}
\begin{split}
	& \frac{d}{d\eta} \delta_{l,\textrm{ctr}} +a \theta_{l,\textrm{ctr}} = 0 \\ & \frac{d}{d\eta} \theta_{l,\textrm{ctr}} + \mathcal{H} \theta_{l,\textrm{ctr}} + \frac{3}{2} \Omega_{\textrm{m,0}} H_{0}^2 \delta_{l,\textrm{ctr}} = ac_{\textrm{eff}}^2 k^2 \delta_{l}^{(1)} \,.
\end{split}
\end{equation}
We can now eliminate the divergence of the velocity field to arrive at
\begin{equation}
\label{eq:systemapp}
	\frac{d^2}{d\eta^2} \delta_{l,\textrm{ctr}} - \frac{3}{2} \Omega_{\textrm{m,0}} H_{0}^2 a(\eta) \delta_{l,\textrm{ctr}} = 2a^{2}(\eta)D_{\textrm{L}}(\eta) c_{\textrm{eff}}^2(\eta) h_{\textrm{ctr}}(\vec{k}) \,, 
\end{equation}
where $\delta_{l}^{(1)}(\eta,\vec{k})=D_{\textrm{L}}(\eta)W_{\Lambda}(k)\delta_{\textrm{L}}(\vec{k})$, and we define
\begin{equation}
\label{eq:hcounter}
	h_{\textrm{ctr}}(\vec{k}) = -\frac{1}{2} k^2 W_{\Lambda}(k)\delta_{\textrm{L}}(\vec{k}) \,.
\end{equation}

In \cite{Nascimento:2024hle} we show that Eq.~(\ref{eq:systemapp}) admits an analytic solution of the form $\delta_{l,\textrm{ctr}}(a,\vec{k}) = c_{\textrm{ctr}}(a)h_{\textrm{ctr}}(\vec{k})$, where:
\begin{equation}
\label{eq:counterapp}
	c_{\textrm{ctr}}(a; \Lambda) = 2H(a) \int_{0}^{a} \frac{da'}{(a')^{3}H^{3}(a')}\int_{0}^{a'} da'' \frac{D_{\textrm{L}}(a'')}{a''} c_{\textrm{eff}}^{2}(a';\Lambda) \,.
\end{equation}
Here we switch from superconformal time to the scale factor and restore the cutoff dependence. The counterterm contribution to the one-loop power spectrum is then
\begin{equation}
\label{eq:powerresapp}
	P_{\textrm{1-loop}}(a,k; \Lambda) =  P_{\textrm{1-loop,SPT}}(a,k; \Lambda) -D_{\textrm{L}}(a)c_{\textrm{ctr}}(a;\Lambda)k^2 W_{\Lambda}^{2}(k) P_{\textrm{L}}(k) \,.
\end{equation}
The 1-loop power spectrum in SPT has the following structure
\begin{equation}
\label{eq:1-loop_spt}
	P_{\textrm{1-loop,SPT}}(a,k; \Lambda) = D_{\textrm{L}}^2(a) W_{\Lambda}^{2}(k) P_{\textrm{L}}(k) + P_{22}(a,k; \Lambda) + P_{13}(a,k; \Lambda) \,,
\end{equation}
where
\begin{equation}
\label{eq:p13_soft}
	P_{13}(a,k \to 0; \Lambda) \approx - \frac{61}{315} D_{\textrm{L}}^{4}(a) k^2 W_{\Lambda}^{2}(k) P_{\textrm{L}}(k) \int_{0}^{\infty} \frac{dq}{2\pi^2} W_{\Lambda}^{2}(q) P_{\textrm{L}}(q) \,,
\end{equation}
in the EdS approximation to perturbation theory kernels \cite{Bernardeau:2001qr}. The soft limit of $P_{22}(a,k; \Lambda)$ is subdominant, with a wavenumber scaling  $\sim \mathcal{O}(k^4)$ as $k \to 0$. 

Self-consistency of the EFT requires that the soft limit of the total 1-loop power spectrum in Eq.~(\ref{eq:powerresapp}) has to be cutoff independent, up to an overall multiplicative factor of $W_{\Lambda}^{2}(k)$. This ensures that UV contributions to perturbation theory loop integrals can be absorbed by the EFT counterterm, and implies \footnote{The subdominant UV contributions from $P_{22}(a,k; \Lambda)$ are absorbed by stochastic terms.} 
\begin{equation}
\label{eq:RGflow}
	c_{\textrm{ctr}}^{(3)}(a;\Lambda) = 	c_{\textrm{ctr}}^{(3)}(a;\Lambda=\infty) + \frac{61}{315} D_{\textrm{L}}^3(a) \int_{0}^{\infty} \frac{dq}{2\pi^2} P_{\textrm{L}}(q) \left[1-W_{\Lambda}^{2}(q)\right] \,.
\end{equation}

This is the Renormalization Group (RG) flow, which entirely fixes the counterterm as a function of the cutoff as soon as the renormalized coefficient, $c_{\textrm{ctr}}^{(3)}(a;\Lambda=\infty)$, is specified (see \cite{Rubira:2023vzw, Rubira:2024tea, Nikolis:2024kbx} for recent papers on the RG flow in the context of biased tracers).

For us this is not the end of the story, as we are interested in working at the level of the effective sound speed, which is related to the counterterm via Eq.~(\ref{eq:counterapp}). This equation can be inverted:
\begin{equation}
\label{eq:inversion}
\begin{split}
	& c_{\textrm{eff}}^{2}(a';\Lambda) = \frac{a}{D_{\textrm{L}}(a)} \frac{d}{da} \left\{ a^3H^{3}(a) \frac{d}{da} \left[ \frac{c_{\textrm{ctr}}(a;\Lambda)}{2H(a)}\right] \right\} \\ = &  \frac{a^4 H^2}{2D_{\textrm{L}}} \left\{\frac{d^2 c_{\textrm{ctr}}}{da^2} + \frac{1}{a} \left(3 + \frac{d \log H}{d \log a}\right) \frac{dc_{\textrm{ctr}}}{da} - \frac{1}{a^2}\left[3 \frac{d\log H}{d\log a} + \left(\frac{d\log H}{d\log a}\right)^2 + \frac{a^2}{H} \frac{d^2H}{da^2} \right] c_{\textrm{ctr}} \right\} \,,
\end{split}
\end{equation}
where in the second line we omit time and cutoff dependencies. The substitution of Eq.~(\ref{eq:RGflow}) into Eq.~(\ref{eq:inversion}) yields
\begin{equation}
\label{eq:RGceff}
\begin{split}
	& c_{\textrm{eff}}^{2}(\Lambda) = c_{\textrm{eff}}^{2}(\Lambda = \infty) + \frac{61}{630} a^2 D_{\textrm{L}}^2 H^2 \Bigg\{ 3f\left(3f-1+\frac{d\log f}{d\log a}\right) + 3f\left(3+\frac{d\log H}{d\log a}\right) \\ & - \left[3\frac{d\log H}{d\log a} + \left(\frac{d\log H}{d\log a}\right)^2 + \frac{a^2}{H} \frac{d^2 H}{da^2} \right] \Bigg\} \int_{0}^{\infty} \frac{dq}{2\pi^2} P_{\textrm{L}}(q) \left[1-W_{\Lambda}^2(k) \right] \,.
\end{split}
\end{equation}
This is the RG flow for the effective sound speed, where $f(a)=d\log D_{\textrm{L}}/d\log a$ is the linear growth rate. Note that Eq.~(\ref{eq:RGceff}) was derived while remaining completely agnostic about the time dependence of the renormalized effective sound speed. This improves upon the considerations made in \cite{Carrasco:2012cv} (while remaining consistent with it), which assume a perturbative time dependence for this quantity in order to derive its RG flow.

In the main text we directly compare our semi-analytic estimate for the effective sound speed, based on separate universe methods, with two previous measurements of the same quantity based on matching the EFT to full N-body simulations. In \cite{Carrasco:2012cv} the authors directly quote the value for the renormalized effective sound speed $c_{\textrm{eff}}^{2}(\Lambda = \infty) \approx 0.6 \times 10^{-6}$ at $z=0$, so that Eq.~(\ref{eq:RGceff}) can be applied as is to compute the effective sound speed as a function of the cutoff scale. 

On the other hand, in \cite{Foreman:2015lca} the authors only extract numerical values for the renormalized counterterm $c_{\textrm{ctr}}(\Lambda=\infty)$. It is important to keep in mind that they assume different conventions than the ones adopted here. In \cite{Foreman:2015lca}, Eq.~(\ref{eq:powerresapp}) should be replaced by
\begin{equation}
	\label{eq:powerresapp_2}
	P_{\textrm{1-loop}}(a,k; \Lambda) =  P_{\textrm{1-loop,SPT}}(a,k; \Lambda) -2(2\pi) c_{s}^{2}(a;\Lambda) D_{\textrm{L}}^{2}(a)\frac{k^2}{k_{\textrm{NL}}^2} W_{\Lambda}^{2}(k) P_{\textrm{L}}(k) \,,
\end{equation}
with for instance $c_{s}^{2}(\Lambda \to \infty) \approx 0.53 \ (k_{\textrm{NL}}/(2h\textrm{Mpc}^{-1}))^{2}$ at $z=0$ extracted from N-body simulations. A direct comparison of Eq.~(\ref{eq:powerresapp_2}) with Eq.~(\ref{eq:powerresapp}) allows us to obtain numerical values for $c_{\textrm{ctr}}(\Lambda=\infty)$. Note that \cite{Foreman:2015lca} denote by the sound speed what we have defined as the EFT counterterm, up to overall multiplicative factors. We then need to apply Eq.~(\ref{eq:counterapp}) to obtain numerical values for $c_{\textrm{eff}}^{2}(\Lambda = \infty)$, and this entails making assumptions about the time dependence of the renormalized effective sound speed. 

We assume the time dependence expected of a scaling universe \cite{Pajer:2013jj},  $P(k) \propto k^{n}$, with $n=-1.5$ following \cite{Ivanov:2018lcg}. This is a good approximation in our fiducial cosmology for the range of scales relevant for the EFT \cite{Angulo:2014tfa, Assassi:2015jqa}. This leads to
\begin{equation}
\label{eq:time-dependence}
	c_{\textrm{ctr}}(a,\Lambda = \infty) = c_{\textrm{ctr}}(a=1,\Lambda = \infty) \left[D_{\textrm{L}}(a)\right]^{\alpha}\,,
\end{equation}
with $\alpha= 4/(n+3)$. We checked explicitly that Eq.~(\ref{eq:time-dependence}) is in good agreement with the results of Table 2 in \cite{Foreman:2015lca} from matching the EFT to N-body simulations at redshifts $z=0$, $z=1$ and $z=2$. The substitution of Eq.~(\ref{eq:time-dependence}) into Eq.~(\ref{eq:inversion}) yields
\begin{equation}
\label{eq:time-dependence_2}
\begin{split}
	& c_{\textrm{eff}}^{2}(a,\Lambda = \infty) = c_{\textrm{ctr}}(a=1,\Lambda = \infty) \frac{a^4 H^2}{2D_{\textrm{L}}} \frac{D_{\textrm{L}}^{\alpha}}{a^2} \times \\ & \times \left\{ \alpha f\left(\alpha f-1+\frac{d\log f}{d\log a}\right) + \alpha f\left(3+\frac{d\log H}{d\log a}\right) - \left[3\frac{d\log H}{d\log a} + \left(\frac{d\log H}{d\log a}\right)^2 + \frac{a^2}{H} \frac{d^2 H}{da^2} \right] \right\}  \,.
\end{split}
\end{equation}
This can now be combined with Eq.~(\ref{eq:RGceff}) to generate the effective sound speed as a function of the cutoff scale from reference \cite{Foreman:2015lca}, at any given redshift.

\section{Halo model response}
\label{sec:app3}

The expression for the response function in the halo model is
\begin{equation}
	\label{eq:response_hm}
	R^{\textrm{HM}}(a,k) = \frac{\left(1+\frac{26}{21} -\frac{1}{3} \frac{d\log P_{\textrm{L}}}{d\log k}\right) \left[I_{1}^{1}(a,k)\right]^2 D_{\textrm{L}}^2(a)P_{\textrm{L}}(k)+I^{1}_{2}(a,k)}{\left[I_{1}^{1}(a,k)\right]^2 D_{\textrm{L}}^2(a)P_{\textrm{L}}(k) + I^{0}_{2}(a,k)} \,,
\end{equation}
where
\begin{equation}
	\label{eq:aux}
	I^{k}_{m}(a,k) = \int dM \frac{dn}{dM}\Big|_{a,M} \left(\frac{M}{\bar{\rho}_{0}}\right)^{m} [b(a,M)]^{k} [u(M|k)]^{m} \,.
\end{equation}
Here $dn/dM|_{a,M}$ is the halo mass function, $b(a,M)$ is the halo bias and $u(M|k)$ is the normalized halo density profile in Fourier space. We assume universality and adopt the halo mass function from Appendix C in \cite{Tinker:2008ff}, halo bias from \cite{tinker2010large} and a Navarro-Frenk-White (NFW) halo profile \cite{Navarro:1995iw}. 

Note that the first line of Eq.~(2.33) in \cite{Wagner:2015gva} contains a typo, since it misses the denominator in Eq.~(\ref{eq:response_hm}) that is just the halo model expression for the power spectrum.

\bibliography{eft.bib}
\bibliographystyle{unsrt}
\end{document}